\documentclass[11pt]{JHEP3}

\usepackage{epsfig}
\newcommand{\ra}{\rightarrow}
\newcommand{\cA}{{\cal A}}
\newcommand{\cB}{{\cal B}}
\newcommand{\cC}{{\cal C}}
\newcommand{\cD}{{\cal D}}

\newcommand{\cJ}{{\cal J}}
\newcommand{\cO}{{\cal O}}

\newcommand{\cL}{{\cal L}}
\newcommand{\cK}{{\cal K}}
\newcommand{\cZ}{{\cal Z}}
\newcommand{\cW}{{\cal W}}

\newcommand{\cT}{{\cal T}}
\newcommand{\Li}{\mbox{Li}}

\newcommand{\bR}{{\bf R}}
\newcommand{\bZ}{{\bf Z}}
\newcommand{\undn}{{\underline n}}
\newcounter{oldcounter}
\addtocounter{equation}{1}
\renewcommand{\baselinestretch}{1.16}

\title{Casimir Energies for 6D Supergravities Compactified
on $\cT_2/Z_N$ with Wilson Lines}

\author{D.M. Ghilencea,${}^1$ D. Hoover,${}^2$
C.P. Burgess${}^{2,3,4}$ and F. Quevedo${}^1$\\

$^1$ DAMTP, Centre for Mathematical Sciences,\\
\qquad Wilberforce Road, Cambridge University, Cambridge CB3 0WA, UK\\
${}^2$ Physics Department, McGill University,\\
\qquad 3600 University Street,
 Montr{\'e}al, Qu{\'e}bec, Canada, H3A 2T8. \\
${}^3$ Department of Physics and Astronomy, McMaster University,\\
\qquad 1280 Main Street West, Hamilton, Ontario, Canada, L8S 4M1.\\
${}^4$ Perimeter Institute,\\ \qquad 31 Caroline Street North,
Waterloo, Ontario, Canada. }

\date{}

\abstract{ We compute (as functions of the  shape and Wilson-line
moduli) the one-loop Casimir energy induced by higher-dimensional
supergravities  compactified from 6D to 4D on 2-tori, and on some
of their $Z_N$ orbifolds. Detailed calculations are
given for a 6D scalar field having an arbitrary 6D mass $m$, and
we show how to extend these results to higher-spin fields for
supersymmetric 6D theories. Particular attention is paid to
regularization issues and to the identification of the divergences of 
the potential, as well as the dependence of the result on $m$, including 
limits for which $m^2 \cA\ll 1$ and $m^2 \cA\gg 1$  where $\cA$ is the 
volume of the internal 2 dimensions. Our calculation extends those in the
literature to very general boundary conditions for fields about
the various cycles of these geometries. 
 The results have potential applications towards Supersymmetric 
Large Extra Dimensions (SLED) as a theory of the Dark Energy. 
First, they provide an explicit
calculation within which to follow the dependence of the result on
the mass of the bulk states which travel within the loop, and for
heavy masses these results bear out the more general analysis of
the UV-sensitivity obtained using heat-kernel methods. Second,
because the potentials we find describe the dynamics of the
classical flat directions of these compactifications, within SLED
they would describe the present-day
dynamics of the Dark Energy.}

\preprint{DAMTP-2004-67, McGill-04/23.}


\begin{document}

\section{Introduction.}

The Casimir energy for various field theories compactified on 2
internal dimensions has been extensively studied (see for example
refs. \cite{Candelas:ae}-\cite{Haba:2002py}). In this paper we
return to this calculation for the particularly simple examples of
2-tori and their orbifolds. At the technical level, our aim
in  is to provide a generalization of previous
calculations for fields compactified on $\cT_2$ and $\cT_2/Z_N$ to
include very general boundary conditions, including those which
would be generated by (constant) Wilson-line backgrounds.

Our physical motivation for performing this calculation is due to
its relevance for the recent proposal for using 6D supergravity
to shed light on the cosmological-constant problem
\cite{SLEDrefs,SLEDreviews} within the context of Supersymmetric
Large Extra Dimensions (SLED). (See also \cite{SLEDrelated} for
related discussions  and similar proposals.) Within this
proposal the extra dimensions must presently be sub-millimeter in
size, and the recently-discovered \cite{DEdiscovery} cosmological
Dark Energy density corresponds to the Casimir energy of the
model's bulk fields as functions of the moduli of these large 2
internal dimensions. The smallness of the Dark Energy density in
this picture is ultimately traced to the large size of these extra
dimensions. In this picture these moduli can be {\it
extremely} light in a technically natural way, and so the Dark
Energy phenomenology is controlled by the dynamics of the moduli
as they respond to this Casimir energy. Although simple
cosmologies appear to be possible along roughly these lines
\cite{ABRS2}, a more detailed determination of their viability
requires a more careful calculation of the modulus potential which
is provided by the Casimir energy, such as we present here.

Torus-based Casimir energies are particularly well-studied, and
recent one-loop studies include \cite{Ito:2003tc}, who study the
energy of massless scalars compactified on $\cT_2$ and $\cT_2/Z_2$
at the special modular point corresponding to an orthogonal
underlying torus. Ref.~\cite{Ponton:2001hq}, on the other hand,
computes the full modulus-dependence of the Casimir energy for
various massless fields compactified on $\cT_2$, assuming these
fields to satisfy periodic boundary conditions about the cycles of
the background geometry. Ref.~\cite{Matsuda:2004ci} computes for
$\cT_2$ compactifications the dependence of the Casimir energy on
a particular modulus, arg$\,U$, but restrict some others (by
choosing equal toroidal radii).
Refs.~\cite{Hosotani:2004ka} and \cite{Hosotani:2004wv} consider
the $\cT_2/Z_2$ orbifold, with moduli fixed to those values
appropriate for an orthogonal underlying torus.
Ref.~\cite{Antoniadis:2001cv} considers the case of $\cT_2/Z_2$
or $\cT_2$, including the presence of Wilson lines, but only
computes the Wilson-line dependent part of the result. Other
recent calculations make similar assumptions
\cite{Hetrick:1989jk,Lee:2000rc, Albrecht:2001cp}.

In this paper we present results for the fields which
appear in supergravity models compactified on a 2-torus, $\cT_2$
and some of its orbifolds, $\cT_2/Z_N$, as functions of the
relevant moduli and the higher-dimensional particle mass, $m$. We
do so for a very general set of boundary conditions about the
cycles of the background geometry, such as could be generated by
the presence of nontrivial Wilson lines wrapping these cycles. We
provide formulae which lend themselves to numerical evaluation,
and focus in particular on the form of the result in the large-
and small-$m$ limits.

A spin-off of this calculation is the information it provides
about the large-$m$ limit, and so of the ultraviolet sensitivity
of these Casimir-energy calculations. This UV-sensitivity is a
crucial part of the SLED proposal, and explicit calculations such
as those presented here provide important checks on the more
general, heat-kernel, UV-sensitivity calculations for 6D
backgrounds, presented in a companion paper, \cite{Doug}.
These general heat-kernel results properly describe the explicit
dependence of the toroidal example considered here, and also show
how flat geometries like tori are dangerous gedanken laboratories,
because they are particularly insensitive to large-$m$ UV effects.

The explicit orbifold calculation  also provides the simplest
example of the appearance of new, brane-localized, UV divergences
--- a phenomenon which is generic to Casimir-energy calculations
in the presence of branes due to the singularities and boundaries
which these branes typically induce in the background geometry
\cite{gilkeycones,Doug}. We exhibit this new divergent
contribution explicitly and  show that it arises  due to the
orbifold projections  which must be  performed.

So far as they go, our results support the SLED framework inasmuch
as all of the one-loop contributions we find are at most of order
$1/\cA^2$, where $\cA$ is the compactification area.
This makes them no larger than is required for the
success of the SLED proposal.

The paper is organized as follows: In the next section we
describe the relevant parts of the toroidal geometry and the
boundary conditions for fields on this geometry for which we
compute the Casimir energy. The explicit calculations are first
performed in detail for complex scalar fields, for which we also
display the ultraviolet-divergent and large-$m$ dependent
parts. Results for the finite part of the Casimir energy are
given for general boundary conditions for massless scalar fields
and these are then extended using a simple argument to obtain the
corresponding results for massless higher-spin fields in 6
dimensions.  Section \ref{z2orbifold} addresses the same
issues for the case of $\cT_2/Z_N$ orbifolds, with the $\cT_2/Z_2$, 
$\cT_2/Z_4$
orbifolds treated in  detail. The small-$m$/large-$m$ cases are
again discussed.  The main text only quotes
the results of the calculations, and full technical
details are provided in the Appendix. Many of the tools of this
Appendix  including the detailed calculations of series
of  Kaluza-Klein integrals,   can prove useful for other
applications such as loop corrections to gauge couplings in gauge
theories on orbifolds with discrete Wilson lines.

\section{Casimir Energy for 2-Tori.}\label{wilson}

In this section we compute the Casimir energy for various 6D
fields in a compactification to 4 dimensions on a 2-torus $\cT_2$, for a
very broad class of boundary conditions for these fields about the
two cycles of $\cT_2$. Although the result is interesting in
its own right, it is also the starting point for the later
calculation of Casimir energy on orbifolds (which exhibit the
effects for the Casimir energy of the presence of co-dimension 2
branes).

We define $\cT_2$  by identifying points on
the plane according to
\begin{equation}\label{rr1}
    (y_1,y_2) \cong (y_1+ n_2 L_2 \cos\theta+n_1 L_1; y_2+ n_2 L_2
    \sin\theta) \,,
\end{equation}
where $n_{1,2}$ are integers and $\theta$, $L_1$ and $L_2$ are the
three real moduli of the torus (see Fig.~\ref{fig1}).
 Equivalently, in terms of the
complex coordinate $z = y_1 + i y_2$ this is
\begin{equation}\label{rr}
    z \cong z +(n_2 U+n_1) L_1 \,,
\end{equation}
with the complex quantity $U$ defined by $U=\exp(i\theta)
L_2/L_1=U_1+i U_2$, $(U_2>0)$.
\FIGURE[ht]{
\centerline{\psfig{figure=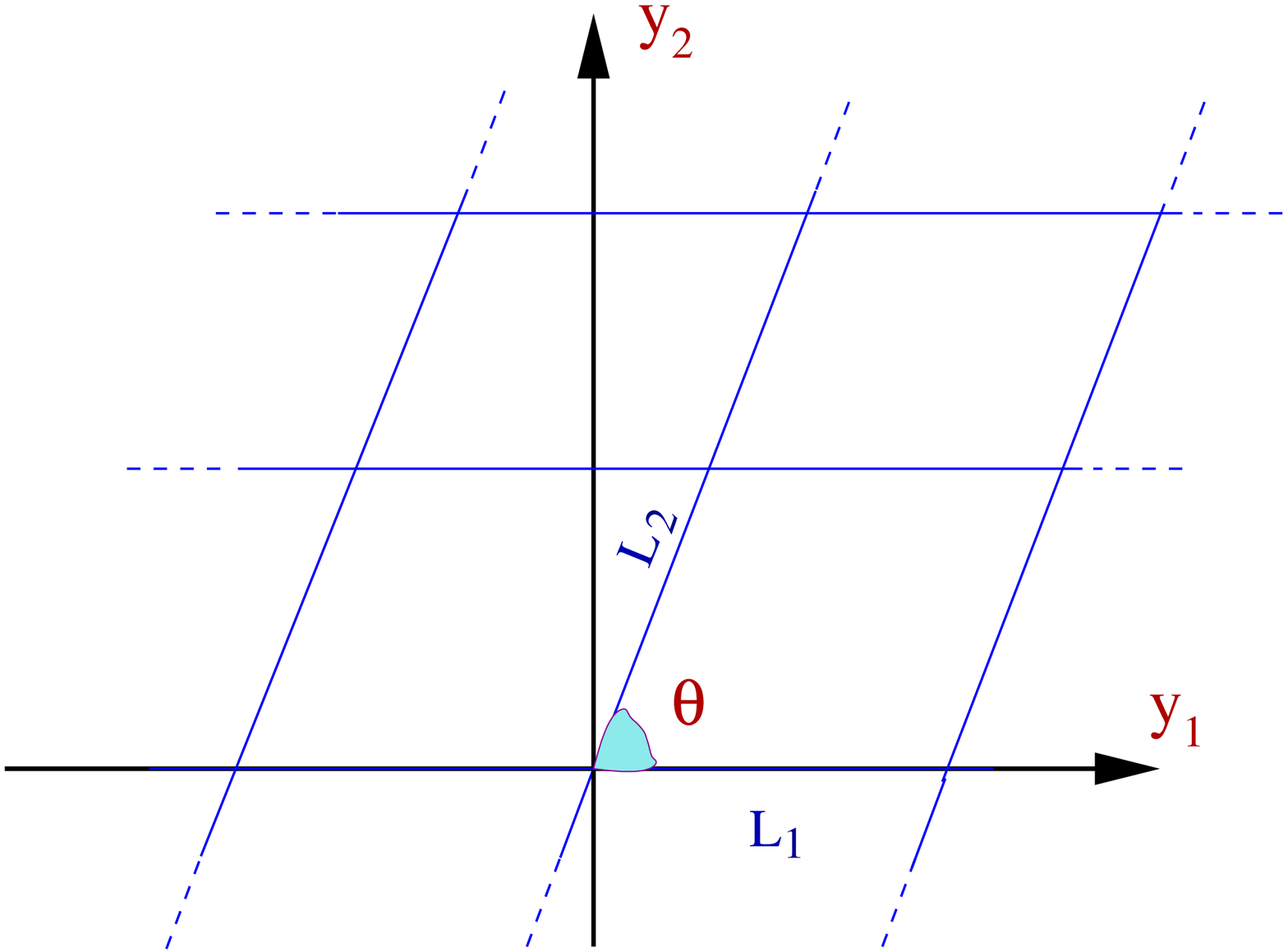,height=1.9in,width=2.6in,angle=0}}
\def\baselinestretch{1.1}
\caption{\small 2D torus characterized by the moduli $\theta$,
$L_1$ and $L_2$, defined by the two identifications $(y_1,y_2) =
(y_1+L_1,y_2)$ and $(y_1,y_2) = (y_1+L_2 \cos\theta, y_2+ L_2
\sin\theta)$.} \label{fig1}}

\subsection{Scalar Field Casimir Energy} \label{torus}

Consider  a complex 6D field $\Phi$ on a space-time compactified
to 4 dimensions on the above 2-torus. Writing the six coordinates
as $\{x_\mu,y_i\}$, with $\mu = 0,...,3$ and $i=1,2$, assume the
scalar satisfies the following boundary conditions
\begin{eqnarray} \label{bcnd}
    \Phi(x,y_1 + n_2 L_2 \cos\theta+n_1 L_1; y_2+ n_2 L_2 \sin\theta)=
    e^{2\pi i(n_1 \,\rho_{1} +n_2 \,\rho_{2})}\,
    \Phi(x,y_1,y_2) \,,
\end{eqnarray}
with $\rho_{1,2}$ being two real quantities. The choices $\rho_{1,2} =
0,\frac12$ correspond to  periodic or anti-periodic
boundary conditions along the torus' two cycles. More general
values of $\rho_i$ are also possible, such as when $\Phi$
transforms non-trivially under a gauge group for which nonzero
Wilson lines are turned on for the corresponding
cycles.\footnote{To see this (for an orthogonal torus)
 notice that the toroidal boundary
conditions preclude removing a constant gauge potential, such as
$A_1 = a$, using only strictly periodic gauge transformations.
(This corresponds to the Wilson line $W = \int_0^{L_1} A_m \, dy^m =
aL_1$.) However, $A_1$ can be removed using a  singular gauge
transformation having parameter $\omega = a\, y_1$,  at
the expense of changing the boundary conditions of charged fields:
$\Phi(y_1 + L_1) = e^{iq a L_1} \, \Phi(y_1)$, where $q$ is
$\Phi$'s charge. We see from this that $2\pi \rho_1 = q a L_1 = q W$.}
Here we consider  $\rho_{1,2}$ arbitrary.

Expanding the scalar field in terms of eigenfunctions
of the 2D Laplacian, $\Box_2 = \partial_1^2 +
\partial_2^2$, according to
\begin{equation} \label{t2modeexpn}
    \Phi(x,y) = \sum_{n_1,n_2} \phi_{n_1,n_2}(x) \,
    f_{n_1,n_2}(y_1,y_2; \rho_1,\rho_2) \,,
\end{equation}
we have $\Box_2 f_{n_1,n_2} = - M_{n_1,n_2}^2 \, f_{n_1,n_2}$ with
\begin{eqnarray}\label{mass22}
    M^2_{n_1,n_2}(\rho_1,\rho_2)&=&
    \frac{(2 \pi)^2}{\cA\, U_2} \, \vert n_2+\rho_{2}-U
    (n_1+\rho_{1})\vert^2 \,.
\end{eqnarray}
Here $\cA\!=\!L_1 L_2 \sin\theta$ denotes the area of the torus.
The mode functions are given by
\begin{equation}\label{wavef}
    f_{n_1,n_2}(y_1,y_2; \rho_1,\rho_2) = \frac{1}{\sqrt \cA}\,
    e^{2 i \, \pi [
    (n_1+\rho_1) (y_1-y_2 \cot\theta)/L_1+ (n_2+\rho_2) y_2/(L_2
    \sin\theta) ]} \,.
    \end{equation}
We now compute the vacuum energy for a complex scalar field having
6D mass $m$, compactified on $\cT_2$ with the above boundary
conditions. Denoting the vacuum-energy per unit 3-volume for such
a scalar field by $V(\rho_1,\rho_2)$, we have
\begin{eqnarray}\label{2torus}
    V(\rho_1,\rho_2) &\equiv& \mu^{4-d}\!\!  \sum_{n_{1,2}\in\bZ}
    \int \!\frac{d^dp}{(2\pi)^d}
    \ln\bigg[ \frac{p^2\! +\! M_{n_1,n_2}^2\! + m^2}{\mu^2} \bigg]\!
    \nonumber \\
\nonumber \\
    &=&
-\frac{\mu^{4}}{(2\pi)^{d}}\!\!\sum_{n_{1,2}\in\bZ}
\int_0^\infty \!\!\!\!\!
\frac{dt}{t^{1+d/2}} e^{-\pi\,t\, \big[\,(M_{n_1,n_2}^2+m^2)/\mu^2\big]}
\end{eqnarray}
where we have continued the momentum integration to Euclidean
signature, and we regulate the ultraviolet divergences which arise
in the sum and integral using dimensional regularization, with the
complex quantity\footnote{The conventions here  differ from those used
in our  previous ref.~\cite{Doug} where  $d = 4-2\,\epsilon$.}
$d = 4 - \epsilon$ ultimately being
taken to 4. Here $\mu$ denotes the arbitrary mass scale which
arises in dimensional regularization, and which drops out of all
physical quantities.

A potential subtlety arises in the above expression for massless
6D fields ($m = 0$) if both $\rho_{1}$ or $\rho_{2}$ are integers,
because in this case there is a choice of integers $(n_1,n_2)$ for
which $M_{n_1,n_2}$ vanishes. In this case it is convenient to
keep $m$ nonzero so that all manipulations remain well-defined,
with $m$ taken to zero at the end of the calculation. In
principle, one must be alive to the possibility of unexpected
singularities appearing when $m$ tends to zero after
renormalization, such as the familiar infrared mass singularities
of Quantum Electrodynamics \cite{RGgood}. As usual, such infrared
problems are less severe in higher dimensions and we shall see
that there is no such obstruction to taking $m \to 0$ for our
applications in 6 dimensions.

The calculation of the two infinite sums in (\ref{2torus}) is
tedious, and is given in detail in Appendices \ref{appendixA} and
\ref{appendix2} --- {\it c.f.} eqs.(\ref{eq1}), (\ref{ldef}),
(\ref{a16}) to (\ref{tz}) --- using the approach of
\cite{Ghilencea:2003kt}. In what follows we quote only the final
results which are appropriate to the discussion at hand. The next
three sections respectively concentrate on the
ultraviolet-divergent part of the result, as well as the finite
part in the cases where the 6D scalars are either massless ($m^2
\cA \to 0$) or very massive ($m^2 \cA \gg 1$).

\subsubsection{Ultraviolet Divergences for 2-Tori}\label{uvdivs}

The ultraviolet divergent part of  $V$ in (\ref{2torus}) denoted
$V_\infty$  is, with $\epsilon=4-d$ (see eqs. (\ref{ldef}),
(\ref{a16}) to (\ref{tz}))
\begin{equation}\label{uvdivergent}
    V_\infty(\rho_1,\rho_2) =
    \frac{m^6 \, \cA}{192 \pi^3 \epsilon}
\end{equation}
which is valid for arbitrary $m$. (Eq.(\ref{uvdivergent}) shows
the importance of keeping a non-zero $m$ when discussing
ultraviolet divergences in dimensional regularization (DR), since
these can easily be missed if $m=0$.) This expression has several
features on which we now remark (and which agree with the more
general analysis of the ultraviolet divergences in 6D field
theories compactified on Ricci-flat backgrounds given in a
companion paper \cite{Doug}).

\bigskip
\noindent $\bullet$ First, the divergent part depends on the
moduli, $L_{1,2}$ and $\theta$, only through the toroidal area
$\cA = L_1 L_2 \sin\theta$, and is interpreted as being a
renormalization of the 6D cosmological constant. Note also that
the UV divergence vanishes if we take $\theta \rightarrow 0$,
corresponding to collapsing the two cycles of the torus onto each
other down to one dimension less.\footnote{This limit must be
treated with care, however, since the calculations of Appendix
\ref{appendixA} also require $U_2\sim \sin\theta$ to be finite and
nonzero.} This agrees with the well-known absence of one-loop UV
divergences for a broad class of theories when they are
dimensionally regularized in odd dimensions.

\smallskip
\noindent $\bullet$ The proportionality to $m^6$ is also what is
required on dimensional grounds (in dimensional regularization)
for a contribution to the 6D cosmological constant. The absence of
other powers of $m$, such as an $\cA$-independent result
proportional to $m^4$, is a consequence of the torus being flat,
and is not true for more general curved spacetimes \cite{Doug}.
Once the UV divergence is renormalized into the 6D cosmological
constant there is no obstruction to taking $m \to 0$, unlike the
situation for massless 4D theories. We consequently feel free to
simply set $m = 0$ in subsequent applications of our formulae for
$V^{\rm ren}$ where
\begin{eqnarray}\label{vren-def}
V^{\rm ren} \equiv V - V_\infty
\end{eqnarray}
with $V$ as in (\ref{2torus}) and $V_\infty$ its divergent part.

\medskip
\noindent$\bullet$ The divergent part of $V$ is
$\rho_i$-independent and so does not depend on the boundary
conditions of the scalar field, eqs.(\ref{bcnd}). Again this agrees
with general arguments, since the
short-wavelength modes responsible for the UV properties are not
sensitive to the boundary conditions which depend on the global
properties of the background geometry. We shall see that for
orbifolds new divergences are present corresponding to
counterterms localized at the fixed points, and these
new divergences can depend on the nature of the boundary
conditions  imposed on the covering space.

\medskip
\noindent$\bullet$ The $1/\epsilon$ pole which appears here
represents a {\it bona fide} 6D divergence. This is at first sight
surprising, since $\epsilon$ represents the difference between $d$
and 4 rather than 6, and it is introduced for each of the
Kaluza-Klein (KK) modes. To understand this it is important to
recognize that our expressions contain two separate sources of UV
divergence: the integration over 4-momentum, $p$, and the two sums over
KK mode numbers, $n_i$. In our calculations the dimensional
continuation is $\epsilon = 4-d$ away from four in order to
regularize the $p$-integration. On the other hand, the KK mode sums
are managed using zeta-function techniques, and the presence of
$\epsilon$ ensures the regularisation of these sums.
With these choices the leading inverse powers of $\epsilon$
obtained turn out to be precisely those which would be obtained
starting from 6D and following the powers of $\epsilon' = (6-d)$.
This equivalence is shown in more detail for an explicit example
(using a spherical geometry, for which more divergences may be
followed) in ref.~\cite{Doug}.

We now examine the finite parts of the Casimir energy density.

\subsubsection{Massless Fields in 6D}

In the massless limit, $m \to 0$, the result for the vacuum energy
density is (see eqs.(\ref{vj}), (\ref{ldef}), (\ref{a16}) to (\ref{tz}))

\begin{eqnarray}\label{wmd}
&& V^{\rm ren}(\rho_1,\rho_2) \vert_{m=0}=
    - \frac{1}{\cA^2}\,
    \left\{\,
    \frac{(2\pi U_2)^3}{90}
    \,\left[\frac{1}{21}-\Delta_{\rho_1}^2 \, \left(1-5\,
    \Delta_{\rho_1}^2-2\,\Delta_{\rho_1}^4+6\,
    \Delta_{\rho_1}^3\right)\right]\right.
\nonumber\\
\nonumber\\
&&\qquad +
    \sum_{n_{1}\in\bZ}\left[
    (n_1+\Delta_{\rho_1})^2\,
    \Li_3(\sigma_{n_1}) + \frac{3\left| n_1 +
    \Delta_{\rho_1}\right|}{2\pi U_2}\,
     \Li_4(\sigma_{n_1}) \right.
    \left. \left.+\frac{3}{(2\pi U_2)^2}\,
    \Li_5(\sigma_{n_1})
    +c.c.\right]\!\right\}
    \nonumber
\end{eqnarray}
with
\begin{eqnarray}
    \sigma_{n_1\geq 0}= e^{-2 i \,\pi \big( \Delta_{\rho_2}-
    U \Delta_{\rho_1} - U n_1\big)}\qquad\hbox{and}\qquad
    \sigma_{n_1<0}=\frac{1}{\overline \sigma_{n_1 > 0}} \,,
\end{eqnarray}
where $\overline \sigma_{n_1}$ denotes the complex conjugate of
$\sigma_{n_1}$. In these expressions $0 \leq\Delta_{\rho_i} < 1$
represents the fractional part of $\rho_i$, as in $\rho_i =
[\rho_i] + \Delta_{\rho_i}$, where $[\rho_i] \in \bZ$ is the
largest integer smaller than or equal to $\rho_i$. The
poly-logarithm functions which appear here are defined by the
sums~\cite{gr}
\begin{equation}
    \Li_\sigma(x) = \sum_{n=1}^\infty \frac{x^n}{n^\sigma} \,.
\end{equation}
\noindent
The point of rewriting the initial two sums into the ones written
here is that these converge well and so are useful for numerical
purposes. Figure \ref{fig2} plots $V(\rho_1,\rho_2)$ as functions
of the moduli $U_1$ and $U_2$ for various choices for the boundary
conditions $(\rho_1,\rho_2)$.

We have checked that the above formula agrees with the particular
cases  studied  in the literature. For instance in the special case
$\rho_{1} = \rho_2 = 0$ we find
\begin{eqnarray}
     V^{\rm ren}(0,0)  \big\vert_{m=0} &=& -\frac{1}{\cA^2} \, \bigg\{
    \frac{4\pi^3 U_2^3}{945}
    + \frac{3 \,\zeta[5]}{2\pi^2 U_2^2}
    +2  \sum_{n_1 =1}^\infty \Big[ n_1^2\, \Li_3(q^{n_1})
    \phantom{\frac12}
     \nonumber \\
\nonumber \\
    && \qquad\qquad  +
    \frac{3\, n_1}{2\pi U_2}\,\Li_4(q^{n_1}) +\frac{3}{4\pi^2 U_2^2}
    \,  \Li_5(q^{n_1})+c.c.\Big]\bigg\},
\end{eqnarray}
where $q\equiv e^{2 i \pi U}$. This agrees with the result given
in ref.~\cite{Ponton:2001hq}.

\subsubsection{Heavy-Mass Dependence} \label{torushmd}

The generality of the calculation in Appendix \ref{appendixA},
\ref{appendix2} also allows the
explicit exhibition of the heavy-mass limit, $m^2 \cA \to \infty$,
of the Casimir energy. This is of particular interest for
Supersymmetric Large Extra Dimensions, where the naturalness of
the description of the Dark Energy density relies on the Casimir
energy only depending weakly on the masses of heavy fields in the
6D bulk.
Using formulae (\ref{a17}), (\ref{jps}), (\ref{tz})
of the appendix it may be shown that
if $m\gg\{ 1/L_1, 1/(L_2 \sin\theta)\}$, leading to $m^2 \cA\gg 1$,
then
\begin{equation}
    V(\rho_1,\rho_2) = \frac{m^6 \, \cA}{384 \pi^3}\Big(\frac{2}{\epsilon}
-\ln\frac{4\pi^3 m^2  e^{\gamma-11/6}}{\mu^2}\Big)
    + \frac{1}{\cA^2} \, \cO\Big(
(m^2  \cA)^\mu \,U_2^\nu \, e^{-2\pi (m^2 \cA)^\sigma
      U_2^{\pm\sigma}}\Big)
\end{equation}
thus powers of $m^2 \cA$ other than $m^6$ are exponentially suppressed.
These expressions agree well with the general results of
ref.~\cite{Doug}, which identify the large-$m$ behaviour using
general heat-kernel techniques. For general geometries there can
be powers of $m$ in the large-$m$ limit, but the leading such
powers are proportional to local effective interactions which
involve polynomials of the background fields and their
derivatives. For the simple toroidal geometries considered here
all of these local interactions vanish, leading to the exponential
mass suppression found above.

The absence of powers like $m^4$ or $m^2$ in the large-$m$ limit
is more difficult to understand from the point of view where the
6D calculation is regarded as simply being the sum over an
infinite number of 4D contributions, each of which can themselves
have such powers of $m$. As is clear from the general 6D analysis
of \cite{Doug}, the absence of these terms may be traced to the
requirements of locality and general covariance in 6 dimensions
--- requirements which are easily missed in a KK mode sum
calculation.

\subsection{Higher-Spin Fields on $\cT_2$.} \label{torushispin}

With an eye towards applications to supersymmetric theories, in
this section we compute the corresponding results for the Casimir
energy for other massless fields in 6 dimensions. We do so using
the trick of ref.~\cite{Ponton:2001hq}, which uses the prior
knowledge that the Casimir energy must vanish once summed over the
field content of a 6D supermultiplet, provided that these fields
all share the same boundary conditions about the cycles of $\cT_2$
(and so do not break any of the supersymmetries).

\TABLE[tbh]{\renewcommand{\arraystretch}{0.95}
\begin{tabular}{|c|c|}
\hline Multiplet & Field Content \\
\hline\hline
 Hyper  &  ($\psi_+$, 2$\phi$)  \\
 Gauge  &  ($A_M$, 2$\psi_-$) \\
 Tensor  & ($A_{MN}^+$, 2$\psi_+$,$\phi$) \\
 Gravitino I& ($\psi_{M+}$, 2$A_{MN}^-$)\\
 Gravitino II &  ($\psi_{M-}$, 2$A_M$,$\psi_-$) \\
 Graviton &  ($g_{MN}$, 2$\psi_{M+}$, $A_{MN}^-$) \\
\hline
\end{tabular}
\caption{The field content of some of the massless representations
of $(2,0)$ supersymmetry in 6 dimensions. The fermions are taken
to be symplectic-Weyl, and their plus and minus subscripts
correspond to their chirality, as measured by their $\Gamma_7$
eigenvalue. The $+$ ($-$) sign on the 2-form potential similarly
indicates it is self (anti-self) dual. There is also an equivalent
set of representations where all plus and minus signs are
interchanged. } \label{masslessReps}}

To this end we reproduce as Table \ref{masslessReps} a table from
ref.~\cite{Doug} listing the massless field content of some of the
representations of $(2,0)$ supersymmetry in 6 dimensions. In this
table the scalars are real, the spinors are symplectic-Weyl and
the 2-form gauge potentials are self-dual or anti-self-dual.

The argument of ref.~\cite{Ponton:2001hq} uses  the
observation that a single symplectic-Weyl fermion and two real
scalars preserve 6-dimensional $(2,0)$ supersymmetry in a toroidal
compactification for which they share the same boundary
conditions, $(\rho_1,\rho_2)$, about the torus' two cycles. The
Casimir energy for these fields must therefore cancel in order to
give a vanishing result for the contribution of a hypermultiplet.
Since we know the scalar result for general $\rho_i$, we may infer
from this that the Casimir energy for a single symplectic-Weyl
fermion must be precisely $-1$ times the result quoted above for a
complex scalar field having the same boundary conditions.

Using the identical argument based on the vanishing of the Casimir
energy summed over the field content of a gauge multiplet
similarly shows that the Casimir energy of a 6D gauge boson must
be $-2$ times the result for a 6D symplectic-Weyl fermion --- and
so is $+2$ times the result for a 6D complex scalar --- having the
same boundary conditions. Arguing in this way allows the inference
of the Casimir energy for all of the other fields appearing in the
supermultiplets listed in Table \ref{masslessReps}. The results
found in this way are
\begin{eqnarray} \label{hispin}
    V_{1/2}(\rho_1,\rho_2) &=& - V(\rho_1,\rho_2) \nonumber\\
    V_{1}(\rho_1,\rho_2) &=& 2\, V(\rho_1,\rho_2) \nonumber\\
    V_{KR}(\rho_1,\rho_2) &=& \frac32 \, V(\rho_1,\rho_2) \nonumber \\
    V_{3/2}(\rho_1,\rho_2) &=& -3\, V(\rho_1,\rho_2) \nonumber\\
    V_{2}(\rho_1,\rho_2) &=& - V(\rho_1,\rho_2) \,,
\end{eqnarray}
where the divergent and finite parts of the right-hand side of
this equation are given explicitly by eqs.~(\ref{uvdivergent}) and
(\ref{wmd}), above. Here $V_{1/2}$, $V_1$, $V_{KR}$, $V_{3/2}$ and
$V_2$ respectively denote the results for a symplectic-Weyl
fermion, a gauge boson, a Kalb-Ramond self-dual (or
anti-self-dual) 2-form gauge potential, a symplectic-Weyl
gravitino and a graviton.

Given these expressions, it is simple to compute the nonzero
Casimir energy which results when 6D supersymmetry is broken {\it
\`a la} Scherk and Schwarz \cite{scherkschwarz}, by assigning
different boundary conditions to different fields within a single
supermultiplet. Supersymmetry is broken in this case by the
boundary conditions themselves. For instance, if the
symplectic-Weyl fermion in a hypermultiplet has boundary condition
$(\rho_1,\rho_2)$ but the complex scalar has boundary condition
$(\rho'_1,\rho'_2)$ then the Casimir energy for this
hypermultiplet would be
\begin{equation}
    V_{\rm hyper} = V_{1/2}(\rho_1,\rho_2) + V(\rho'_1,\rho'_2)
    = V(\rho'_1,\rho'_2) - V(\rho_1,\rho_2) \,.
\end{equation}
Similarly applying different boundary conditions to the
constituents of a gauge or tensor multiplet gives
\begin{eqnarray}
    V_{\rm gauge} &=& V_1(\rho_1,\rho_2) + 2 \,
    V_{1/2}(\rho'_1,\rho'_2) = 2\, [ V(\rho_1,\rho_2) -
    V(\rho'_1,\rho'_2)] \nonumber\\
    V_{\rm tensor} &=& V_{KR}(\rho_1,\rho_2)
    + 2 \,V_{1/2}(\rho'_1,\rho'_2) + \frac12 \, V(\rho''_1,\rho''_2)
    \nonumber\\
    &=& \frac32 \, V(\rho_1,\rho_2) -2 \, V(\rho'_1,\rho'_2)
    + \frac12 \, V(\rho''_1,\rho''_2) \,,
\end{eqnarray}
and so on.

 Figure \ref{fig2} gives in addition to
the  plots of $V(\rho_1,\rho_2)$,
 the differences $V(\rho_1,\rho_2) -
V(\rho'_1,\rho'_2)$ as functions of $U_1$ and $U_2$ for various
choices for the boundary conditions $(\rho_1,\rho_2)$ and
$(\rho'_1,\rho'_2)$. The periodicity wrt $U_1$ in the plots is a
remnant of the $SL(2,Z)_U$ symmetry (modified by non-zero Wilson
lines). For $U_2=L_2/L_1 \sin\theta\geq \cO(1)$, one has flat
directions for $V$ (as function of $U_1,U_2$). This changes for
$U_2\ll 1$ (say if $\theta\ll 1$) when $V$ develops maxima/minima.
For values of $\rho_i$ other than those in the figure, the peaks
in these plots have different height.

\FIGURE[t]{
\begin{tabular}{cc|cc|}
\parbox{7.1cm}{
\psfig{figure=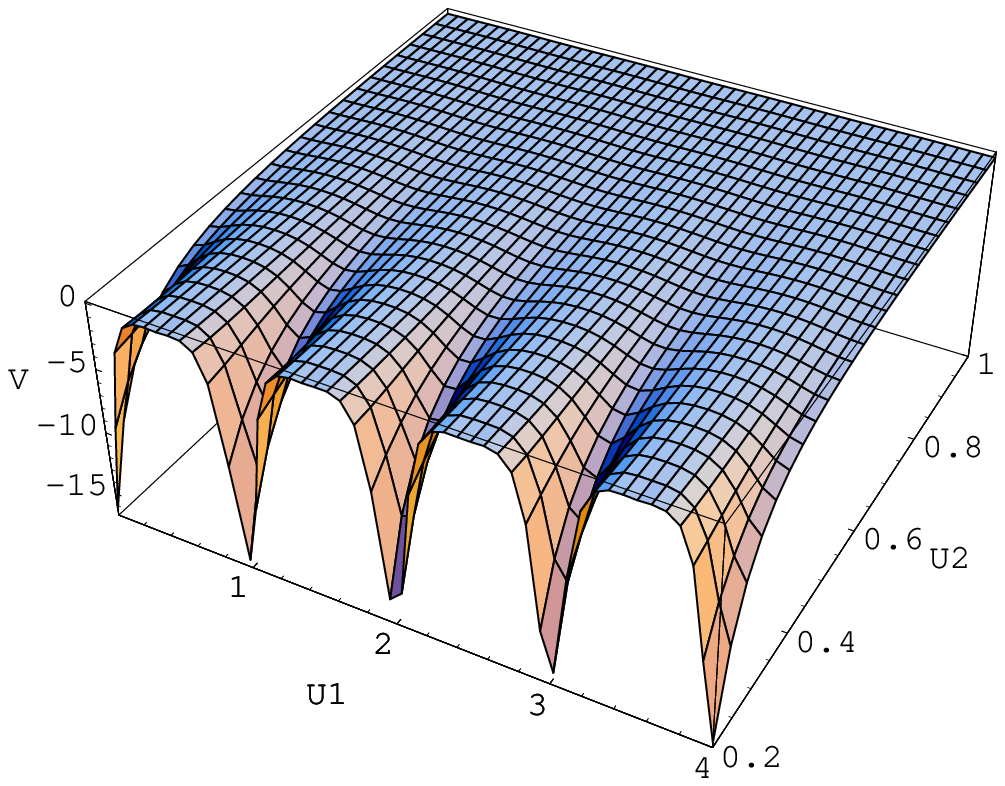,height=5.4cm,width=5.66cm}} \hfill{\,\,}
\parbox{7.1cm}{
\psfig{figure=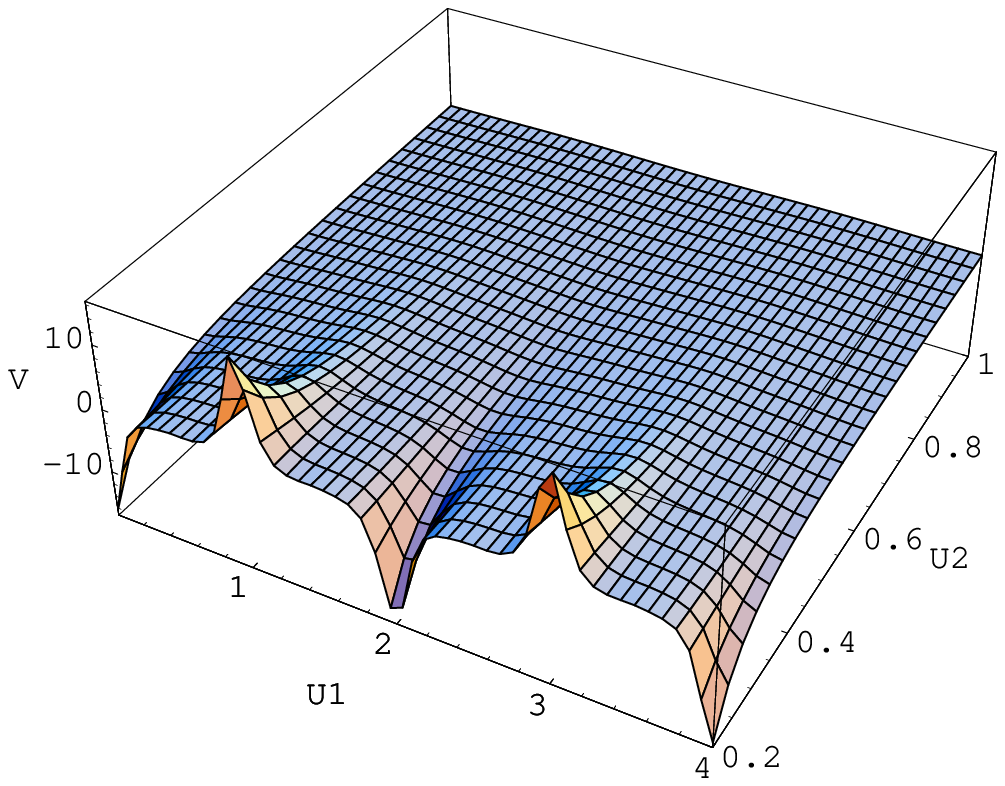,height=5.4cm,width=5.66cm}}
\end{tabular}
   \begin{tabular}{cc|cc|}
   \parbox{7.1cm}{\small (a) The potential $V^{ren}$  for $\rho_1=\!\rho_2=0$.\\}
   \hfill{\,\,}
   \parbox{7.1cm}{\small (b) The potential $V^{ren}$  
for $\rho_1=\!1/2,\,\rho_2\!=0$.\\}
   \end{tabular}
$  $
\begin{tabular}{cc|cc|}
\parbox{7.1cm}{
\psfig{figure=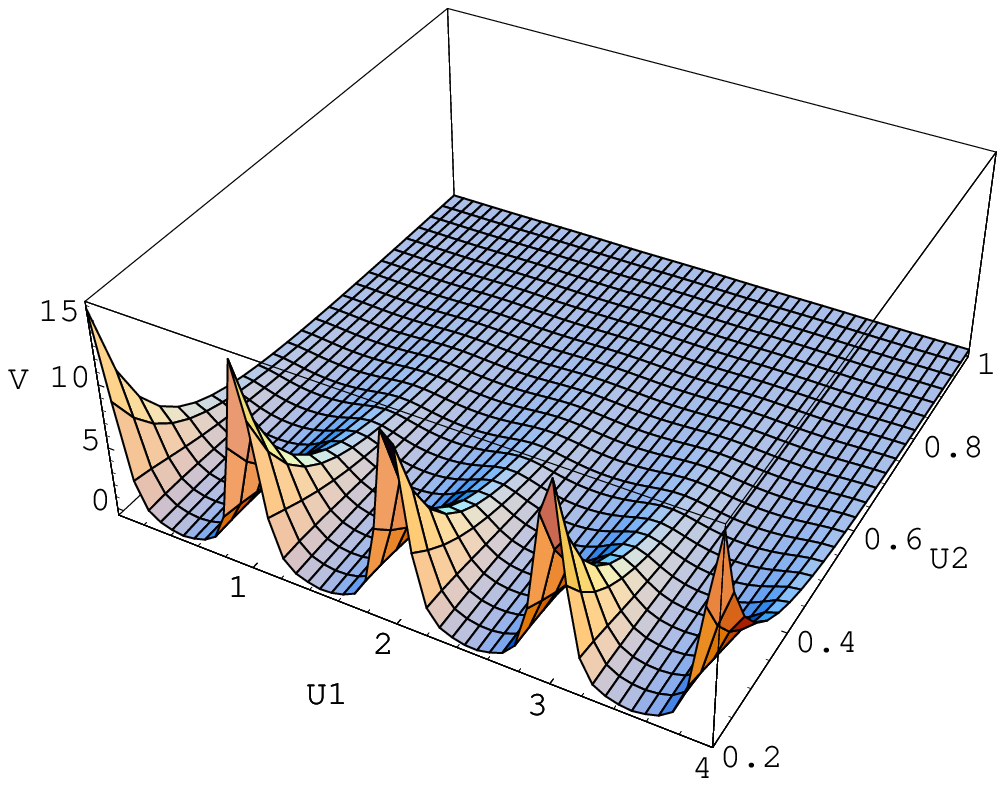,height=5.4cm,width=5.66cm}}
\hfill{\,\,}
\parbox{7.1cm}{
\psfig{figure=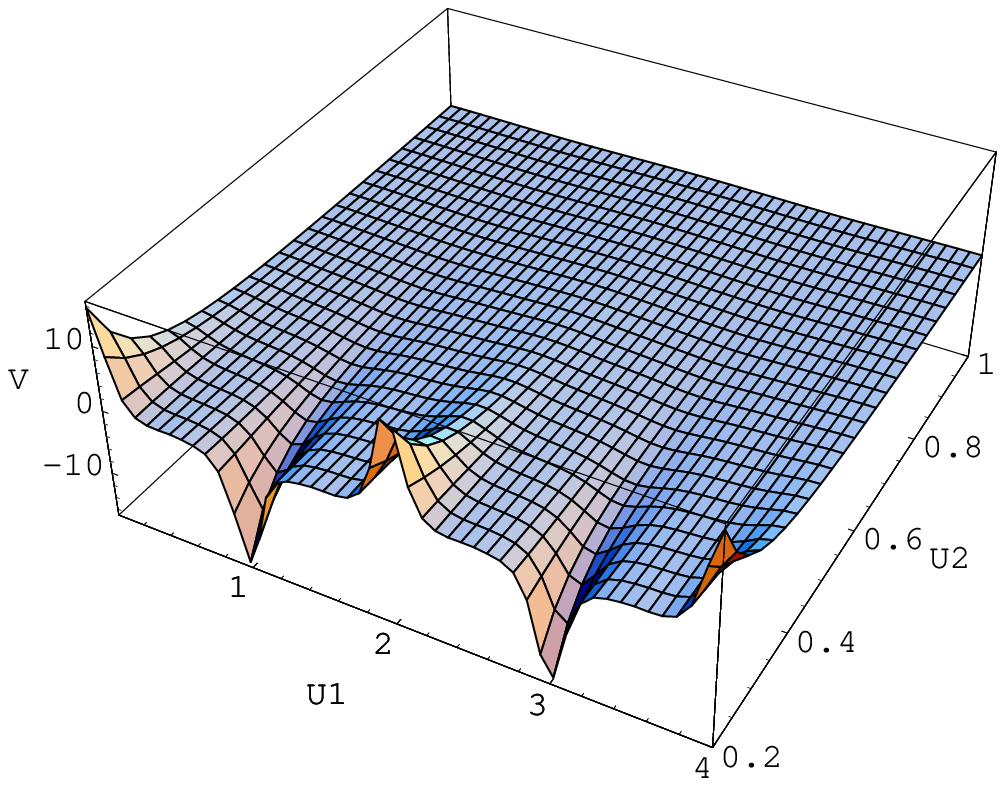,height=5.4cm,width=5.66cm}}
\end{tabular}
    \begin{tabular}{cc|cc|}
    \parbox{7.1cm}{\small (c) The potential $V^{ren}$  
for $\rho_1=0, \rho_2=1/2$.\\}
    \hfill{\,\,}
    \parbox{7.1cm}{\small (d) The potential $V^{ren}$
for $\rho_1\!=\!1/2,\rho_2\!=\!1/2$.\\}
    \end{tabular}
\begin{tabular}{cc|cc|}
\parbox{7.1cm}{
\psfig{figure=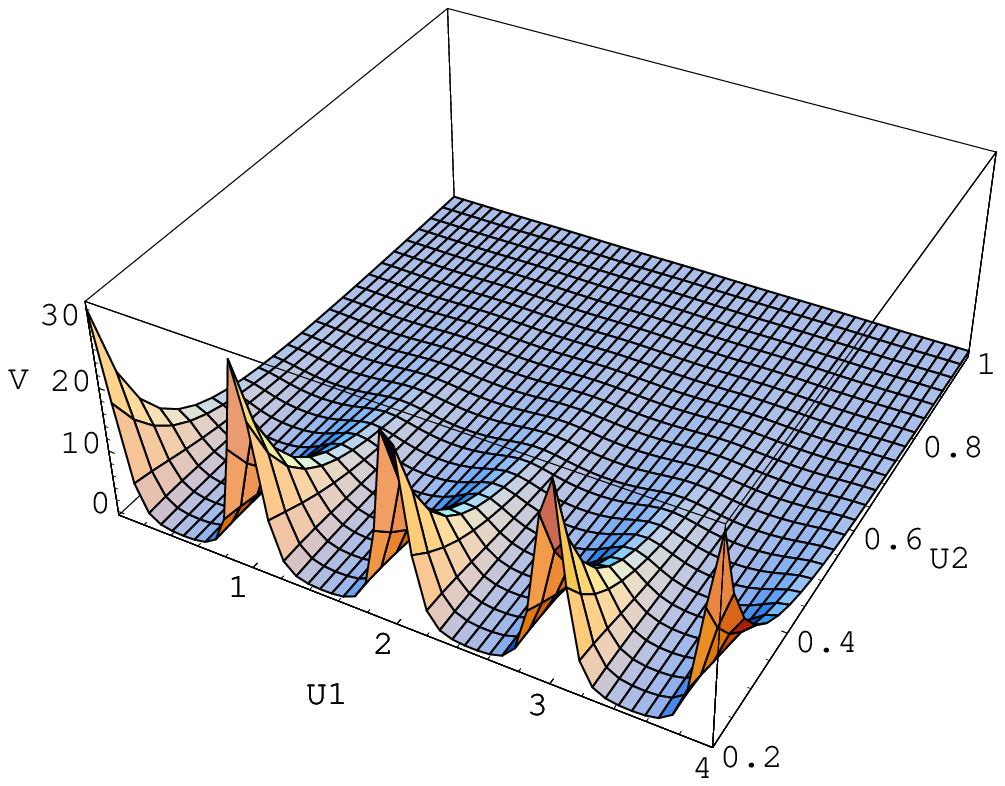,height=5.4cm,width=5.66cm}}
\hfill{\,\,}
\parbox{7.1cm}{
\psfig{figure=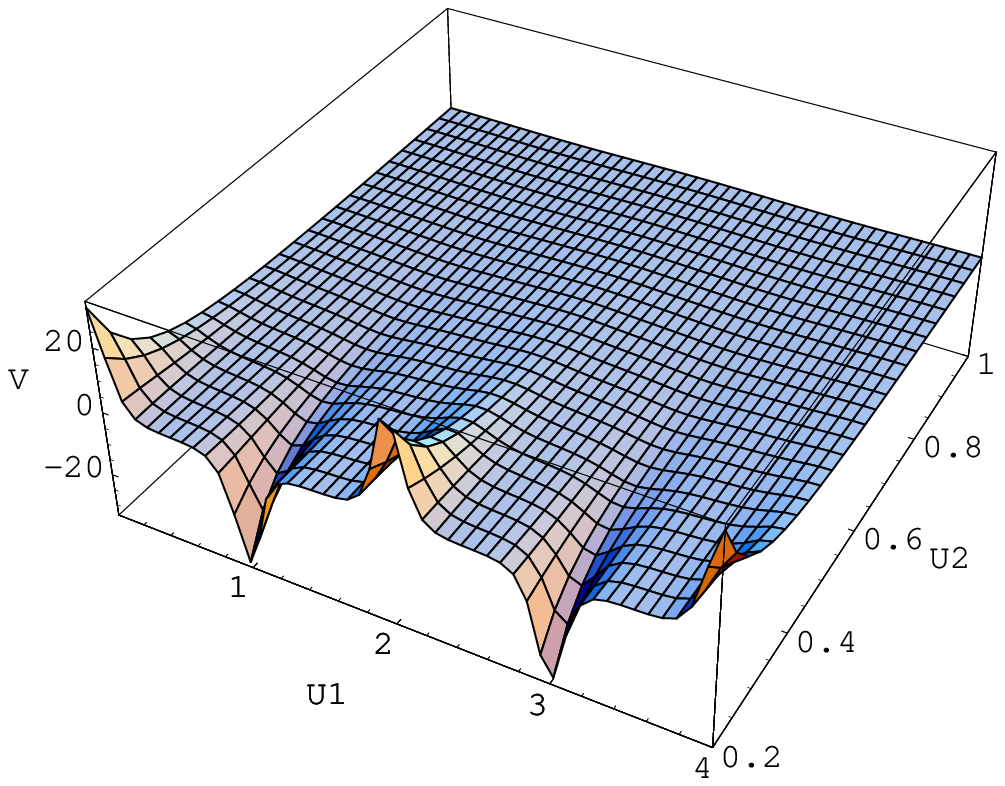,height=5.4cm,width=5.66cm}}
\end{tabular}
   \begin{tabular}{cc|cc|}
   \parbox{7.1cm}{\small (e) The plot
$V^{ren}\big\vert_{\rho_1\!=\!0,\,\rho_2\!=\!\frac{1}{2}}
-V^{ren}\big\vert_{\rho_1=\!\rho_2\!=\!0}$}
   \hfill{\,\,}
   \parbox{7.1cm}{\small (f) The plot
$V^{ren}\big\vert_{\rho_1\!=\!\rho_2\!=\!\frac{1}{2}}-
V^{ren}\big\vert_{\rho_1=\frac{1}{2},\,\rho_2=0}$}
   \end{tabular}
\vspace{0.6cm}
\caption{
\small{The (finite part of the) potential $V$,
up to an overall {\it positive} factor and for fixed  area $\cA$.
 The plots are in function of
$U_1$, $U_2$, ($U\!\equiv\!U_1\!+\!i U_2\!=\!L_2/L_1 \exp(i \theta)$)
for various  $\rho_{1,2}$. See also eqs.(\ref{wmd}) and
(\ref{VV}). The peaks of the plots indicate (moduli) divergences present at
 $U_2\ll 1$ when the two dimensions collapse onto each other
 ($\theta=0$). }}
\label{fig2}}
\cleardoublepage

\section{Casimir Energy for Orbifolds}
\label{z2orbifold}

None of the previous results included the effects of 3-branes
within the bulk when computing the Casimir energies. We now extend
these calculations to some simple examples which include branes,
and for which the background geometry includes the back-reaction
of the brane tensions by incorporating the appropriate conical
singularities at the brane positions. We only consider here brane
singularities which correspond to the specific defect angles which
arise when an orbifold is constructed from the 2-torus by
identifying points under the action, $Z_N$, of a discrete set of
rotations. We analyze separately the case of  $\cT_2/Z_2$, with
full  details, and then the orbifold $\cT_2/Z_{4}$
whose technical details  differ considerably. For  $\cT_2/Z_{4}$
we use a general  method which can be  applied to
the remaining $\cT_2/Z_3$ and $\cT_2/Z_6$.

To this end we return to the description of $\cT_2$ as a complex
plane, $z= y_1+i \, y_2$, identified under the action of a lattice
of discrete translations as in eq.~(\ref{rr}). Following standard
practice, we construct an orbifold from this torus by further
identifying points under the action of the $Z_N$ rotations defined
by
\begin{equation}
    z \cong \tau^k \, z \,,
\end{equation}
where $\tau = e^{2 \pi i/N}$. This gives a well-defined coset
space, ${\cal O}= \cT_2/Z_N$, provided that these rotations
take the initial lattice which defines the torus onto itself.
Notice that if $N=2$, then the rotation $z \cong - z$ is
automatically a symmetry of the lattice for any value of the
moduli $L_1$, $L_2$ and $\theta$, and so these three quantities
are also moduli of the resulting orbifold, ${\cal O}$. On
the other hand, if $N>2$ then the rotation is a symmetry of the
lattice only for specific choices for the complex structure: $U
\equiv (L_2/L_1)e^{i\theta} = \tau$, and so $L_2 = L_1 = L$ and
$\theta = 2 \pi/N$, and so only  one modulus, $L$, in this case
survives.

The coset $\cT_2/Z_N$ is an orbifold rather than a manifold
because of the metric singularities which arise at the fixed
points of the group. For instance, in the case $\cT_2/Z_2$ there
are 4 such points, corresponding to $z = 0, \frac12,
\frac{1}{2} \, U$ and $\frac12(1 + U)$. The metric has a conical
singularity at each of these points, whose defect angle is  $\pi$.
For further details on orbifolds and their fixed points see
Appendix~\ref{appendixF}.

\subsection{A Scalar Field on $\cT_2/Z_2$.} \label{z2orbifoldspin0}

We now compute the Casimir energy for a complex scalar field,
$\Phi$, compactified from 6D to 4D on the orbifold ${\cal O}
= \cT_2/Z_2$. As before, the 6 coordinates are taken to be
$\{x,y_i\}$, with the orbifold corresponding to the
coordinates $y_i, i = 1,2$. We consider the 6D scalar field to
satisfy the  boundary conditions
\begin{eqnarray}\label{bc1}
    \Phi(x,y_1+ L_1; y_2) & =&  e^{2\pi i \rho_1}\,\, \Phi(x,y_1,y_2);
    \qquad \rho_1=0,1/2.
    \\    \Phi(x,y_1+ L_2 \cos\theta; y_2+ L_2 \sin\theta)& =&
    e^{2\pi i  \rho_2}\,\, \Phi(x,y_1,y_2);\qquad \rho_2=0,1/2.\label{bc2}
    \\
    \Phi(x,-y_1,-y_2)& = &\pm\,\, \Phi(x,y_1,y_2) \,.\label{parity}
\end{eqnarray}
Condition (\ref{parity}) is possible because
of the new cycle that the orbifold has (which the torus does not).
As is indicated in eqs.~(\ref{bc1}) and (\ref{bc2}), the
quantities $\rho_i$ are no longer free to take any real value in
this case, because the underlying Wilson lines must be compatible
with the orbifold rotation. For instance, when acting on the
coordinates it is straightforward to show that the composite
transformation $X = \Sigma \circ P \circ \Sigma$ gives $X(y_1,y_2) =
(-y_1,-y_2)$, if $\Sigma(y_1,y_2) = (y_1 + L_1,y_2)$ and $P(y_1,y_2) =
(-y_1,-y_2)$. Applying the same transformations to $\Phi$ and
using the boundary conditions (\ref{bc1}) and (\ref{parity}),
consistency requires $\exp(4\pi i \rho_1) = 1$, and so $\rho_1 =
n/2$ for some integer $n$. A similar argument implies that
$\rho_2$ is also half-integer. If we require, without loss of
generality, $0 \le \rho_i < 1$, then we see that consistency
requires $\rho_{1,2}=0,1/2$. Altogether there are 8 possible
choices for the boundary conditions, denoted
by $(\rho_1,\rho_2)^{\pm}$.

Because the orbifold reflection is a symmetry of $\Box_2$, these
reflections have a natural action on its eigenfunctions,
$f_{n_1,n_2}$. Using the explicit expressions obtained earlier,
eq.~(\ref{wavef}), for the toroidal mode functions
\begin{eqnarray}\label{wavefaction}
    f_{n_1,n_2}(-y_1,-y_2; \rho_1,\rho_2) & = & \frac{1}{\sqrt \cA}\,
    e^{-2 i \, \pi [
    (n_1+\rho_1) (y_1-y_2 \cot\theta)/L_1+ (n_2+\rho_2) y_2/(L_2
    \sin\theta) ]}
\nonumber\\
    &=&
f_{n_1,n_2}^*(y_1,y_2; \rho_1,\rho_2) \,,
\nonumber\\
&    \equiv& f_{n_1',n_2'}(y_1,y_2; \rho_1,\rho_2) \,,
\end{eqnarray}
where $n'_i = -n_i - 2 \rho_i$ ensures $n'_i + \rho_i = -(n_i +
\rho_i)$. Notice that $n'_i$ defined in this way remains an
integer because for the $Z_2$ orbifold $\rho_i = 0,\frac12$.

Using this action it is straightforward to specialize the toroidal
mode expansion, eq.~(\ref{t2modeexpn}), to fields on $\cT_2/Z_2$
with boundary conditions $(\rho_1,\rho_2)^{\pm}$. We now write
these expansions explicitly, in terms of the real
and imaginary parts of the mode functions $f_{n_1,n_2} = e_{n_1,n_2} + i
g_{n_1,n_2}$.

\bigskip\noindent{\bf (1).} {$(\rho_1,\rho_2)=(0,0)^\pm$}. In this
case $f_{n_1,n_2}(-y_1,-y_2) = f_{-n_1,-n_2}(y_1, y_2)$ and so using the
boundary conditions of eqs.~(\ref{parity}) in the mode expansion
of eq.~(\ref{t2modeexpn}) gives
\begin{eqnarray}\label{par1}
    \Phi^+(x,y_1,y_2) & =& \sum_{n_1\geq 0,\,n_2\in\bZ}
    \frac{2}{2^{\delta_{n_1,0}}}
    \,\phi_{n_1,n_2}(x) \,\, e_{n_1,n_2}(y_1,y_2;0,0),
    \nonumber\\
    \Phi^-(x,y_1,y_2) &= & \sum_{n_1\geq 0,\, n_2\in\bZ}'
    \frac{2i}{2^{\delta_{n_1,0}}}
    \, \phi_{n_1,n_2}(x) \, \,g_{n_1,n_2}(y_1,y_2;0,0),
\end{eqnarray}
where the superscript on $\Phi$ indicates the sign chosen for the
orbifold projection, and $\delta_{n_1,0}$ is the usual Kronecker
delta-function which vanishes unless $n_1 = 0$, in which case it
equals unity. Notice that because $g_{0,0}(y_1,y_2; 0,0)=0$,
the mode $(n_1,n_2)=(0,0)$ is absent in the double sum
for $\Phi^-$, as is indicated by the primed double sum.

\bigskip\noindent {\bf (2).} $(\rho_1,\rho_2)=(0,1/2)^\pm$.
In this case $f_{n_1,n_2}(-y_1,-y_2) = f_{-n_1,-n_2-1}(y_1,y_2)$ and so the
mode expansion becomes
\begin{eqnarray}\label{par2}
    \Phi^+(x,y_1,y_2) & =& \sum_{n_1\geq 0,\,n_2\in\bZ}
    \frac{2}{2^{\delta_{n_1,0}}}
    \,\phi_{n_1,n_2}(x) \,\, e_{n_1,n_2}(y_1,y_2;0,1/2),
    \nonumber\\
    \Phi^-(x,y_1,y_2) &= & \sum_{n_1\geq 0,\, n_2\in\bZ}
    \frac{2i}{2^{\delta_{n_1,0}}}
    \, \phi_{n_1,n_2}(x) \, \, g_{n_1,n_2}(y_1,y_2;0,1/2) .
\end{eqnarray}

\bigskip\noindent
{\bf (3).} $(\rho_1,\rho_2)=(1/2,0)^\pm$. Here $f_{n_1,n_2}(-y_1,-y_2) =
f_{-n_1-1,-n_2}(y_1,y_2)$, and so
\begin{eqnarray}\label{par3}
    \Phi^+(x,y_1,y_2) & =& \sum_{n_1\geq 0,\, n_2\in\bZ}
    2 \,\,\phi_{n_1,n_2}(x) \,\, e_{n_1,n_2}(y_1,y_2;1/2,0),\qquad
    \nonumber\\
    \Phi^-(x,y_1,y_2) &= & \sum_{n_1\geq 0,\,n_2\in\bZ}
    2 i\, \phi_{n_1,n_2}(x) \,\, g_{n_1,n_2}(y_1,y_2;1/2,0),\qquad
\end{eqnarray}

\bigskip\noindent
{\bf (4).}  $(\rho_1,\rho_2)=(1/2,1/2)^\pm$. In this case
$f_{n_1,n_2}(-y_1,-y_2) = f_{-n_1-1,-n_2-1}(y_1,y_2)$, and so
\begin{eqnarray}\label{par4}
    \Phi^+(x,y_1,y_2) & =& \sum_{n_1\geq 0,\, n_2\in\bZ}
    2\,\,\phi_{n_1,n_2}(x) \,\, e_{n_1,n_2}(y_1,y_2;1/2,1/2),\,\,
    \nonumber\\
    \Phi^-(x,y_1,y_2) &= & \sum_{n_1\geq 0,\, n_2\in\bZ}
    2i\, \phi_{n_1,n_2}(x) \,\, g_{n_1,n_2}(y_1,y_2;1/2,1/2) .\,\,
\end{eqnarray}
There are two ways to
 compute the Casimir energy for the scalar field $\Phi$ on
$\cT_2/Z_2$ with these boundary conditions.
One approach is to recognize that the scalar
propagator on the orbifold may be obtained from the propagator on
the torus using the method of images:
\begin{equation}
    G^\pm_{{\cal O}}(x-x',y-y') = G_{\cT}(x-x',y-y') \pm
    G_{\cT}(x-x',y+y') \,,
\end{equation}
and following the implications of this for the vacuum energy. The
second approach is to directly perform the KK mode sum over the
modified mode functions given above. Both lead to the same result,
and we present the mode-function derivation here because, albeit more
involved,  it can be  extended to the case of  $\cT_2/Z_N$ and
allows a general discussion of the ultraviolet divergences for $\cT_2/Z_N$.

The vacuum energy density per unit 3-volume written as a mode sum
is given by
\begin{eqnarray}\label{orbifold1}
V_\cO(0,0)^\pm &=&
    \sum_{n_1\geq 0,\,n_2\in\bZ}^{(-)'}
    \frac{\mu^{4-d}}{2^{\delta_{n_1,0}}}
    \int \frac{d^dp}{(2\pi)^d}\,
    \ln\bigg[ \frac{p^2 + M_{n_1,n_2}^2(0,0) + m^2}{\mu^2} \bigg]
  \nonumber\\
  \nonumber\\
V_\cO(0,1/2)^\pm &=&
    \sum_{n_1\geq 0,\,n_2\in\bZ} \frac{\mu^{4-d}}{2^{\delta_{n_1,0}}}
    \int \frac{d^dp}{(2\pi)^d}\,
    \ln\bigg[ \frac{ p^2 + M_{n_1,n_2}^2(0,1/2) + m^2}{\mu^2} \bigg]
 \nonumber\\
 \nonumber\\
V_\cO({1}/{2},0)^\pm &=&
    \sum_{n_1\geq 0,\, n_2\in\bZ} \,\mu^{4-d}\,
    \int \frac{d^dp}{(2\pi)^d}\,
    \ln\bigg[ \frac{p^2 + M_{n_1,n_2}^2(1/2,0) + m^2}{\mu^2} \bigg]
\nonumber\\
\nonumber\\
V_\cO({1}/{2},{1}/{2})^\pm &=&
    \sum_{n_1\geq 0,\, n_2\in\bZ} \,\mu^{4-d}\,
    \int \!\frac{d^dp}{(2\pi)^d}\,
    \ln\bigg[\frac{ p^2 + M_{n_1,n_2}^2(1/2,1/2) + m^2}{\mu^2}
    \bigg] ,
\end{eqnarray}
where $d = (4 - \epsilon)$ and the symbol $(-)'$ on the sum in the
first line indicates the exclusion from the sum of the single mode
$(n_1,n_2) = (0,0)$, but only for the case of $(0,0)^-$ boundary
conditions.

As might be expected from  the approach  based on
the method of images, these expressions may be evaluated in terms
of the corresponding quantities on the torus. To see this for the
mode sums we denote the summands of these expressions by ${\cal
W}(|n_2 + \rho_2 - U(n_1 + \rho_1)|^2)$ in order to emphasize
their dependence on mode numbers and moduli. It is then simple to
use the invariance of ${\cal W}$ under changes in sign of $(n_i +
\rho_i)$ to prove the following identities:
\begin{eqnarray}\label{par}
    \sum_{n_1\geq 0;\, n_2\in\bZ} \frac{1}{2^{\delta_{n_1,0}}}
    \cW\Big( \vert n_2-U n_1\vert^2\Big) & =&
    \frac{1}{2} \sum_{n_1, n_2 \in \bZ}
    \cW\Big( \vert n_2 - U n_1\vert^2\Big)
    \nonumber\\ \nonumber\\
    \sum_{n_1\geq 0;\, n_2 \in\bZ}
    \frac{1}{2^{\delta_{n_1,0}}}
    \cW\Big( \vert n_2+1/2 - U n_1\vert^2\Big)
     &=& \frac{1}{2} \sum_{n_1,n_2\in\bZ}
    \cW\Big( \vert n_2+1/2 - U n_1\vert^2\Big)
    \nonumber\\
\nonumber\\
    \sum_{n_1\geq 0;\, n_2\in\bZ}
    \cW\Big(\vert n_2+\rho_2-U(n_1+1/2)\vert^2\Big)
    & =& \frac{1}{2}\!\! \sum_{n_1, n_2\in\bZ}
    \cW\Big(\vert n_2+\rho_2-U (n_1+1/2)\vert^2\Big),\quad
\end{eqnarray}

\noindent
where $\rho_2$ is either 0 or $\frac12$ in the last line.
These expressions allow the derivation of the following
expressions for the Casimir energies in terms of the toroidal
results, $V(\rho_1,\rho_2)$
\begin{eqnarray}\label{orbifold}
    V_{\cal O}(0,0)^- &=&
    \frac12 \, \Bigl[ V(0,0) -
    V_{\rm zm} \Bigr] \nonumber\\
    V_{\cal O}(\rho_1,\rho_2)^\pm &=& \frac{1}{2}\,
    V(\rho_1,\rho_2) \qquad \hbox{for all others.}
\end{eqnarray}
Here $V_{\rm zm}$ is the contribution to the torus Casimir
energy  of the ``zero mode''  $(n_1, n_2)\!=\!(0,0)$ with
$(\rho_1,\rho_2)^-=(0,0)^-$ (for its expression see (\ref{zmode}) and
(\ref{tz})).
We can now present  in  detail the divergent and finite parts of
the sums and integrals in (\ref{orbifold1}), (\ref{orbifold}).

\subsubsection{Ultraviolet Divergences}
\label{sectuvdivs}

We isolate the divergent part of  $V_\cO$ in
eqs.(\ref{orbifold1}) and write
\begin{eqnarray}
V_\cO(\rho_1,\rho_2)^\pm= V_{\cO, \infty} (\rho_1,\rho_2)^\pm
+V_\cO^{\rm{ren}}(\rho_1,\rho_2)^\pm
\end{eqnarray}
where all divergent terms are included in $V_{\cO,\infty}$. As is
clear from eqs.~(\ref{orbifold}), for all choices of boundary
condition except $(0,0)^-$ on the orbifold the ultraviolet
divergences encountered are precisely half of those encountered on
the torus, eq.~(\ref{uvdivergent}):
\begin{eqnarray}\label{uvdivergentorb}
    V_{{\cal O}\infty}(\rho_1,\rho_2)^\pm =
    \frac{m^6 \, \cA}{384 \pi^3 \epsilon}
    =  \frac{m^6 \, \cA_{\cal O}}{192 \pi^3 \epsilon}
    \,, \qquad \hbox{if} \qquad
    (\rho_1,\rho_2)^\pm \ne (0,0)^-
    \,.
\end{eqnarray}
This divergence may be absorbed, as usual, into a renormalization
of the bulk cosmological constant. Notice that its coefficient is
the same as was obtained earlier for the torus, once the
divergence is expressed in terms of the area of the orbifold,
$\cA_{\cal O}$, which is half the area, $\cA$, of the covering
torus.

By contrast, the exclusion of the zero mode for the specific
choice $(0,0)^-$ introduces a new type of divergence which was not
encountered for the torus. In this case the orbifold and toroidal
divergences differ by the contribution of the $n_1 = n_2 = 0$ mode
alone, and thus, using eqs.(\ref{vj}), (\ref{ldef}), (\ref{a16}),
(\ref{a17}), one has
\begin{eqnarray}
    V_{{\cal O}\infty}(0,0)^- = \frac12 \, V^*_{\infty}(0,0)
    = \frac{m^6 \, \cA_{\cal O}}{192 \pi^3 \epsilon}
    + \frac{m^4}{32 \pi^2 \epsilon} \,.
\end{eqnarray}
The presence of the
 last term is consistent with the general heat-kernel
analysis. Because it is proportional to $m^4$ and is independent
of the bulk moduli, it has the right properties to be interpreted
as a renormalization of the tension of the branes whose presence
at the fixed points is responsible for the conical singularities
in the bulk geometry at these points. As before, the
ultraviolet-finite part of the Casimir energy obtained after this
renormalization is nonsingular in the $m \to 0$ limit, and so we
are free to take this limit explicitly in the renormalized result
for massless 6D fields which we quote below.

It should be emphasized that although this divergence renormalizes
the local brane tensions, it arises due to the functional
integration over {\it bulk} fields. This is a feature which arises
quite generically for quantum effects in the presence of
boundaries and defects, whose origin can be understood in detail
as follows. The bulk vacuum energy {\it density}, $t_{MN} =
\langle T_{MN} \rangle$ is ultraviolet finite (for the flat
orbifold under discussion) after the bulk cosmological constant is
appropriately renormalized. However although $t_{MN}$ is finite,
it is also position-dependent due to the presence of the orbifold
singularities breaking the translation invariance of the
underlying torus. In particular, $t_{MN}(y)$ typically goes to
infinity as the singular points are approached in a way which
diverges once integrated over the volume of the orbifold. It is
this new divergence which is renormalized by the brane-tension
counter-term localized at the singularity.

\subsubsection{6D Massless Fields}

Using eqs.(\ref{par}) we can now  give the explicit
results for the Casimir energy of a massless complex 6D scalar
field compactified on $\cT_2/Z_2$, for the various boundary
conditions $(\rho_1,\rho_2)^\pm$. It is noteworthy that $V^{\rm
ren}_{\rm zm}(0,0)$ vanishes as $m \to 0$ in dimensional
regularization, and so the orbifold result is half of the
appropriate toroidal result for all choices of boundary
conditions. After the renormalization of the ultraviolet
divergences described above, one has
\begin{eqnarray}\label{qwqw}
V_\cO^{\rm ren}(\rho_1,\rho_2)^\pm\big\vert_{m=0}=\frac{1}{2} V^{\rm
  ren}(\rho_1,\rho_2)\big\vert_{m=0}, \qquad \rho_{1,2}=0,1/2.
\end{eqnarray}
and that
\begin{eqnarray} \label{VV}
&&\quad
V^{\rm ren}_{\cal O}(0,0)^\pm \Big\vert_{m=0}
=   -\frac{1}{\cA^2}\, \bigg\{ \bigg( \frac{1}{21} \bigg)
    \frac{(2\pi U_2)^3}{180}
    + \frac{3 \,\zeta[5]}{(2\pi U_2)^2}
\nonumber\\
\nonumber\\
    && \hspace{3.15cm} +\sum_{n_1 = 1}^\infty \Big[
    n_1^2\, \Li_3(q^{n_1}) +\frac{3\, n_1}{2\pi U_2}
    \,\Li_4(q^{n_1}) +\frac{3}{(2\pi U_2)^2} \,  \Li_5(q^{n_1})
    +c.c.\Big]\bigg\}
\nonumber\\
\nonumber\\
\nonumber\\
    && V^{\rm ren}_{\cal O}(0,1/2)^\pm  \Big\vert_{m=0}
 =  -\frac{1}{\cA^2}\, \bigg\{ \bigg( \frac{1}{21} \bigg)
    \frac{(2\pi U_2)^3}{180} -
    \frac{45\,\zeta[5]}{16 (2\pi U_2)^2}
      \nonumber\\
\nonumber\\
    && \hspace{3.15cm}
    +\sum_{n_1 = 1}^\infty
    \Big[ n_1^2\, \Li_3(-q^{n_1}) +\frac{3\,n_1}{2\pi U_2}
    \,\Li_4(-q^{n_1}) +\frac{3}{(2\pi U_2)^2} \,  \Li_5(-q^{n_1})
    +c.c.\Big]\bigg\}
\nonumber\\
\nonumber\\
\nonumber\\
  && V^{\rm ren}_{\cal O}({1}/{2},0)^\pm \Big\vert_{m=0} =
    -\frac{1}{\cA^2}\, \bigg\{ -\bigg(\frac{31}{672} \bigg)
    \frac{(2\pi U_2)^3}{180}
    + \sum_{n_1 = 0}^\infty \Big[\big( n_1 + 1/2 \big)^2 \,
    \Li_3 \big(q^{n_1+{1}/{2}} \big)
\nonumber\\
\nonumber\\
  &&\hspace{3.15cm}+ \,\frac{3\,
    \big(n_1 + 1/2 \big)}{2\pi U_2} \,
    \Li_4 \big(q^{n_1+{1}/{2}} \big)
       + \frac{3}{(2\pi U_2)^2} \,\Li_5 \big(q^{n_1+{1}/{2}} \big)
    + c.c.\Big] \bigg\}
\nonumber\\
\nonumber\\
\nonumber\\
    && V^{\rm ren}_{\cal O}({1}/{2},{1}/{2}) \Big\vert_{m=0} =
    -\frac{1}{\cA^2}\, \bigg\{ -\bigg(\frac{31}{672} \bigg)
    \frac{(2\pi U_2)^3}{180}
 + \sum_{n_1 = 0}^\infty \Big[ \big(n_1 +{1}/{2} \big)^2\,
    \Li_3 \big(-q^{n_1+{1}/{2}} \big)
\nonumber\\
\nonumber\\
    && \hspace{3.15cm}
    + \, \frac{3\, \big(n_1+1/2
    \big)}{2\pi U_2} \Li_4 \big(-q^{n_1 + {1}/{2}} \big)
    + \frac{3}{(2\pi U_2)^2} \, \Li_5\big(-q^{n_1+{1}/{2}} \big)
    \!+\! c.c.\Big]\!\!\bigg\}\qquad\,
\end{eqnarray}
where $q = e^{2\pi i U}$ and the complex conjugate applies only to
the series of polylogarithms.

\subsubsection{Heavy-Mass Dependence} \label{orbifoldhmd}

The divergences of the Casimir energy for large $m$ are identical
to those for the case of small $m$ discussed in Section
\ref{sectuvdivs}. Further, because the orbifold results are simply
expressed in terms of the toroidal ones, the heavy-mass dependence
of the toroidal expressions carry over immediately to the orbifold
Casimir energy. In particular, in dimensional regularization  (and
after modified minimal subtraction) the
finite parts of the Casimir energy fall exponentially for large
$m$, and the only strong $m$-dependence arises in the divergent
terms, including the new $m^4$ term which arises for some of the
boundary conditions.

\subsubsection{Higher-Spin Fields on $\cT_2/Z_2$}
\label{z2orbifoldhispin}

The results for massless higher-spin fields on the $\cT_2/Z_2$
orbifold  can be read  from their toroidal
counterparts of Section~\ref{torushispin}.
This is possible because the orbifold
identification does not break supersymmetry provided that all of
the fields within a 6D supermultiplet satisfy the same boundary
conditions. The results for the Casimir energy of higher-spin
fields may therefore simply be read off by multiplying the
expressions (\ref{VV}) by the factors given in
eqs.~(\ref{hispin}).

\subsection{The Orbifold $\cT_2/Z_N$ with $N>2$. }
\label{znorbifold}

In this section we outline the steps for computing the Casimir
energy for a complex scalar field $\Phi$ compactified on the $\cT_2/Z_N$
orbifolds, with $N>2$. Recall that for these orbifolds
\begin{eqnarray}
U=e^{2 i\pi/N},\qquad L_2=L_1=L,\qquad \theta = 2 \pi/N
\end{eqnarray}
We take the following action of the translation and
orbifold $Z_N$ symmetries on the field $\Phi$
\begin{eqnarray}\label{t2z2}
    \Phi^g(x,\tau^k\, z)&=& g^k \, \Phi^g(x,z),
    \nonumber\\
    \Phi^g(x,z+ L_1) &=& e^{2 i \pi \rho_1}\, \Phi^g (x,z)
    \nonumber\\
    \Phi^g(x,z+U L_1) &=& e^{2 i \pi \rho_2 }\, \Phi^g (x,z) \,,
\end{eqnarray}
where we use complex coordinates $z = y_1 + iy_2$ and as before
$\tau\equiv e^{2 i \pi/N}$ and $k =\overline{ 0,N-1}$. Here $g$ is a
particular representation of the $Z_N$ transformation, acting on
$\Phi^g$. The superscript `$g$' emphasizes that this
representation is not unique, and the explicit form taken by
$\Phi^g$ in general depends on which $g$ is chosen. As in the
case of $\cT_2/Z_2$, this realization only faithfully reproduces the
symmetry for specific choices for the $\rho_i$, whose values we
now determine.

The consistency conditions for the action on $\Phi^g$ are found by
combining the above expressions and using geometrical relations
which state how some of the $Z_N$ rotations can also be expressed
as translations on the covering torus. To display these we use the
complex coordinate $z = y_1 + iy_2$, in terms of which the lattice
of translations which defines the underlying torus is generated by
$e_1 = L_1$ and $e_2 = U L_1$. Then, depending on the group of
rotations, $Z_N$, which is of interest, the following restrictions
can arise.
\begin{itemize}
\item If an orbifold rotation takes $e_1$ to $e_2$ --- {\it i.e.}
there is an integer $0 < k < N$ for which $\tau^k e_1 = e_2$ (or
$\tau^k = U$) --- then using eqs.~(\ref{t2z2}) to evaluate
$\Phi^g (x,\tau^k(z + e_1)) = \Phi^g (x,\tau^k z + e_2\big)$ implies
\begin{equation}
    g^k \, \exp(2 \pi i \rho_1) = g^k \, \exp(2 \pi i \rho_2) \,,
\end{equation}
and so we may take $\rho_1 = \rho_2$ without loss of generality.
\item If an orbifold rotation takes $e_i \to -e_i$ then a similar
argument implies
\begin{equation}
    g^k \, \exp(2 \pi i \rho_i) = g^k \, \exp(-2 \pi i \rho_i) \,,
\end{equation}
and so we may take $\rho_i = 0$ or $1/2$.
\end{itemize}
For instance, only the second of these conditions applied to the
$Z_2$ orbifold considered previously. By contrast, both conditions
apply to the case of $\pi/N$ rotations which give the orbifold
$\cT_2/Z_{2N}$, and so for this case we must take $\rho_1 = \rho_2
= 0,\frac12$. For the $Z_3$ case, on the other hand, the first
condition applies but instead of the second condition one has $e_2
+ \tau^k e_2 = - e_1$, and so we find $\rho_2 = \rho_1 = 0,
\frac13$ or $\frac23$. These results express the quantization on
these orbifolds of the underlying Wilson lines which are
responsible for the boundary conditions which are expressed by the
$\rho_i$. For a more detailed description of Wilson lines and their
values on orbifolds see Appendix~\ref{appendixF}.

To determine the action of the symmetries  (\ref{t2z2}) on the toroidal mode
functions, we adapt the discussion of Section~2.3 of
ref.~\cite{Scrucca:2003ut} to include the general phases
$\rho_{1,2}$. It is convenient for these purposes to rewrite
eq.~(\ref{wavef}) in complex coordinates
\begin{equation}
    f_{\undn_1,\undn_2} (z,\bar z) =\frac{1}{\sqrt \cA} \,\, e^{
    [(\undn_2 -\undn_1 {\overline U}) \, z\, -\,
    (\undn_2 -\undn_1  U) \, {\overline z}]/(2 L_1 U_2) },
\end{equation}
where $\undn_i = n_i + \rho_i$, for $i=1,2$. The construction of
the mode functions for the orbifold $\cT_2/Z_N$ is done by
observing that all of the arguments in \cite{Scrucca:2003ut}
remain valid if $n_i$ is replaced by $\undn_i = n_i + \rho_i$. The
basis functions one is led to are given by
\begin{eqnarray}\label{hf}
    h^g_{\underline n_1,\undn_2}(z)=\frac{1}{\sqrt N}\,\,
    \eta_{{\underline n},{\underline 0}}
    \sum_{k=0}^{N-1}  g^{-k} f_{\underline n_1,\undn_2}(\tau^k \, z)
\end{eqnarray}
where $\eta_{{\underline n},{\underline 0}}=1/{\sqrt N}$ if
${\underline n}_1 = \undn_2 =0$, and otherwise equals 1. Using this
definition, one can check that the mode functions satisfy the
orbifold condition
\begin{eqnarray}
    h^g_{\underline n_1,\undn_2}(\tau^k z)= g^k\,
    h^g_{\underline n_1,\undn_2}(z) \,.
\end{eqnarray}
As outlined in \cite{Scrucca:2003ut} not all the functions
$h_{\undn_1 \undn_2}$ are independent, since they are related by
\begin{eqnarray}\label{ind}
    h_{\omega^k(\tau)\cdot(\undn_1,\undn_2)}^g (z)=
    h^g_{\undn_1,\undn_2} (\tau^k\,z),
\end{eqnarray}
where $\omega^k(\tau)\cdot(\undn_1,\undn_2)$ is defined as a rotation
 which takes $-\undn_1+\tau \undn_2$ into $-\undn_1'+\tau \undn_2'=\tau^k
(-\undn_1+\tau \undn_2)$. The set of all such rotations
({\it i.e.} for $k=0,1,..,N-1$) identify $N$ domains
whose union covers the whole complex plane $(Ox; Oy)$ defined by
$(Ox; Oy)=(-\undn_1+\tau_1 \undn_2; \tau_2 \undn_2)$.
Each such domain fixes the set
of levels $\undn_1,\undn_2$ which identify the set of independent
$h_{\undn_1,\undn_2}^g$ which are not related by the rotation
$\omega$ of (\ref{ind}). This gives
 $\undn_1\! =\! n_1\! +\! \rho_1\! <\! 0$ and $\undn_2\! =\! n_2\! +\!
\rho_2 \geq 0$ as an independent set. Since here $0\!\leq\!
\rho_{1,2}\!<\!1$, one concludes that the conditions $n_1\! <\! 0,
n_2\! \geq\! 0$ define an independent set of functions
$h_{\undn_1,\undn_2}$ for the orbifold $\cT_2/Z_N$.

Using the above considerations, one has the following mode
decomposition for $\Phi^g$:
\begin{eqnarray}
    \Phi^g(x,z) = \sum_{n_1<0,n_2\geq 0}
    \Phi^g_{\underline n_1,\undn_2}(x) \, \, h_{\underline n_1,\undn_2}(z)
    + \frac{\delta^{g,1}}{\sqrt \cA} \,\,
    \Phi^1_{\underline 0,\underline 0} ,
    \qquad \undn_i=n_i+\rho_i \,,
\end{eqnarray}
which satisfies the desired condition $\Phi^g(x,\tau^k z)\! =\! g^k \,
\Phi^g(x,z)$. For example, for $\cT_2/Z_3$
\begin{eqnarray}\label{eqexp}
    g=\tau^0,\qquad \Phi^0(x, z)
    &=& \sum_{n_1<0,n_2\geq 0} \phi^0_{\undn_1,\undn_2}(x)\,
    \,\bigg[\frac{1}{\sqrt 3} \sum_{k=0}^{2}
 f_{\undn_1,\undn_2}(\tau^k \,z)\bigg]
    +\frac{1}{\sqrt \cA} \phi^0_{{\underline 0}, {\underline 0}}(x)
    \nonumber\\
    g=\tau^1,\qquad \Phi^1(x,z)
    &=& \sum_{n_1<0,n_2\geq 0} \phi^1_{\undn_1,\undn_2}(x)\,
    \, \bigg[\frac{1}{\sqrt 3} \sum_{k=0}^{2} \tau^{-k}
    \, f_{\undn_1,\undn_2}(\tau^k \,z)\bigg]
    \nonumber\\
    g=\tau^2,\qquad \Phi^2(x,z)&=& \sum_{n_1<0,n_2\geq 0}
\phi^2_{\undn_1,\undn_2}(x)\,
    \, \bigg[\frac{1}{\sqrt 3} \sum_{k=0}^{2} \tau^{-2 k}\,
f_{\undn_1,\undn_2}(\tau^k \,z)\bigg]
\end{eqnarray}

\vspace{0.2cm} \noindent The main difference from the
$\cT_2/Z_2$ orbifold is that for $\cT_2/Z_N$ the sum over the
Kaluza-Klein levels is restricted to positive/negative values of
$n_{1,2}$, unlike in $\cT_2/Z_2$ where one  sum could be extended
to the whole set $\bZ$ of integers. The above mode expansion leads
to the following expression for the Casimir energy of a complex 6D
scalar field on $\cT_2/Z_N$, $N\!>\!2$.

\begin{eqnarray}\label{vorb}
    V_{\cal O}\left(\rho_1,\rho_2\right) &=&
    \sum_{n_1<0,\,n_2\geq 0} \mu^{4-d}
    \int \frac{d^dp}{(2\pi)^d}
    \ln\bigg[ \frac{p^2 + M_{n_1,n_2}^2(\rho_1,\rho_2) +
    m^2}{\mu^2} \bigg] \,
\end{eqnarray}
\noindent where for each $g$ one uses the values of
$\rho_{1,2}$ which respect the consistency conditions.

The above domain of summation for $n_{1,2}$ makes the analytical
calculation of $V_{\cal O}(\rho_1,\rho_2)$ more difficult than
in the case of $\cT_2/Z_2$. The difficulty is
 caused by the fact that none of the sums over  $n_1$,
$n_2$ can be extended\footnote{with some exceptions in the case of
$\cT_2/Z_4$ orbifolds, see later.} to a sum over the whole set
$\bZ$ of integers (as we had for $\cT_2/Z_2$, eq.(\ref{orbifold1})).
 As a result  no (Poisson)
resummation of individual contributions to $V(\rho_1,\rho_2)$ is
possible and  the calculation is then more tedious.
Although  one may still be able to work on the covering
torus\footnote{For a general approach to computing traces on orbifold
  spaces see  \cite{GrootNibbelink:2003gd}.}
rather than in the orbifold basis, the approach below (being valid
for any $\rho_{1,2}$) allows a simultaneous analysis of all  orbifolds
$\cT_2/Z_N$, $N\!>\!2$.

After a long calculation (see Appendix \ref{appendixC},
eqs.(\ref{sss}) to (\ref{lll})) one has for $V(\rho_1,\rho_2)$
of (\ref{vorb}) the following result (which is valid for $m\ll 1/L$)

\begin{eqnarray}\label{vorb2}
V_\cO(\rho_1,\rho_2)& =& -\frac{1}{2 \,\cA^2}\,
\bigg\{\tilde \cD+
\frac{m^6 \cA^3}{768\,\pi^3}
\frac{-2}{\epsilon}
+\frac{(2\pi  U_2)^3}{180}  \bigg[\frac{1}{21}\!-\rho_1^2
  (1-5\rho_1^2 -\! 2 \rho_1^4+6 \rho_1^3)\bigg]\!
\nonumber\\
\nonumber\\
+&&\!\!\!\!\!
\sum_{n_1<0}
\bigg[ (n_1+\rho_1)^2 \, \Li_3(\sigma_{n_1})
\!+\!\frac{3 \,\vert n_1+\rho_1\vert }{2\pi \,U_2}\, \Li_4(\sigma_{n_1})
\!+\!\frac{3 }{4\pi^2 U_2^2}  \, \Li_5(\sigma_{n_1})\!+\!c.c.
\bigg]\!\bigg\}\qquad\,\,
\\
\nonumber
\end{eqnarray}
where
\begin{eqnarray}\label{sig2}
\sigma_{n_1<0}= e^{-2 i \pi \big( U (n_1+\rho_1)-\rho_2\big)},\qquad
0\leq \rho_i<1,\qquad U=U_1+i\, U_2, \,\,U_2>0.
\end{eqnarray}
This is the result for the Casimir energy for $\cT_2/Z_N$,
$N\!>\!2$ with boundary conditions as in (\ref{t2z2}) and with
$\rho_{1,2}$ taking the values required by the consistency
conditions specific to each orbifold. Finally, $\tilde\cD$ of
(\ref{vorb2}) is an  asymptotic series given by (see
Appendix~\ref{appendixC}, eq.(\ref{dfdf}))

\begin{equation}\label{ddefinition}
\tilde\cD \! = \!
\frac{(\mu^2 \cA)^\frac{\epsilon}{2}}{\pi^{{\epsilon}/{2}-2}}
 \!\!\!\!\!\!
\sum_{n_1<0, p\geq 0}
\!\frac{2\, (-1)^p}{p!\, U_2^{2 p-2}}\,
\Gamma\Big[p-2+\frac{\epsilon}{2}\Big]\,
\zeta\big[-2 p, \rho_2-U_1 (n_1+\rho_1)\big]\,
\Big[\frac{m^2 \cA}{(2\pi)^2 U_2}+(n_1+\rho_1)^2\Big]^{2-p-\frac{\epsilon}{2}}
\end{equation}
where $\zeta[q,x]$ is the Hurwitz zeta function \cite{gr},
$\cA=L^2\sin\theta$, $U_2={\rm Im} U=\sin\theta$, $\theta=2\pi/N$.
This expression of $\tilde\cD$  is valid without any restrictions
on the relative values of $m$, $L$ or $U$.

The quantity $\tilde\cD$ is of particular interest because it
contains additional poles as $\epsilon \to 0$, and so potentially
introduces new contributions to the UV divergent part of $V_\cO$
in (\ref{vorb2}). Note that if one of the sums (say that over
$n_2\geq 0$) in $V_\cO$ of (\ref{vorb}) were extended to the whole
$\bZ$ set of integers, the quantity $\tilde\cD$ given above would
not arise due to the cancellation against the similar contribution
to $V_\cO$, coming from $n_2<0$. The latter would actually be
equal to $\tilde\cD$ of (\ref{ddefinition}) with the substitutions
$\rho_2\ra 1-\rho_2$ and $U_1\ra -U_1$.  The sum of these two
contributions would then vanish
\begin{eqnarray}
\tilde\cD
(\rho_1,\rho_2)+\tilde\cD(\rho_1,1-\rho_2)
\Big\vert_{U_1\ra -U_1}=0
\end{eqnarray}
since $\zeta[-2 m, x]+\zeta[-2 m, 1-x]=0$. In particular, this
explains the absence of $\tilde\cD$ in $\cT_2$ and $\cT_2/Z_2$
where one of the KK sums was over the whole set $\bZ$. To
conclude, the presence of $\tilde\cD$ in the potential $V_\cO$ is
due to the fact that {\it both} sums over the Kaluza-Klein modes
in (\ref{vorb}) were restricted to positive/negative modes only.

\subsubsection{Ultraviolet Divergences}

Because the contribution $\tilde\cD$ potentially introduces new UV
divergences for $\cT_2/Z_N$ orbifolds with $N>2$, in this section
we investigate their form in more detail in order to see what
kinds of counterterms they require. In particular, we show that
for $\cT_2/Z_{2N}$ no new counterterms are required beyond those
which already arise for the $Z_2$ orbifold. To do so we consider
in detail the case of $\cT_2/Z_4$, for which the analysis of
$\tilde\cD$ is considerably simplified. For the remaining cases
$\cT_2/Z_N$ (with $N=3,6$) the analysis follows the same technical
steps as below, but is more involved and  will be presented
elsewhere \cite{dmg}.

In the case of $\cT_2/Z_4$ which has $U_1=0$, $U_2=1$, the Hurwitz
zeta function in (\ref{ddefinition}) has no dependence on $n_1$,
and this simplifies the identification of the additional poles. In
the last bracket in eq.(\ref{ddefinition}) one can then use a
binomial expansion ($m L\ll 1$)  or an asymptotic expansion ($m
L\gg 1$) and following the technical details in Appendix
\ref{appendixC}, eqs.(\ref{dd}), (\ref{rrr}), (\ref{qqq}) one
obtains from (\ref{ddefinition}) that, for $\cT_2/Z_4$

\begin{eqnarray}\label{coeffs}
\!\!\!\!\!\!\tilde\cD\!  &=&- \frac{\cA^2}{8 \pi^2} \bigg\{
\!\frac{m^4}{\epsilon} \,c_1 +\!\frac{2 m^2}{\epsilon}
\frac{(2\pi)^2}{\cA} \, c_2\!+  \frac{1}{\epsilon}\,
\frac{(2\pi)^4}{\cA^2} \, c_3  +\!\tilde\cD_f\!
+\!\cO(\epsilon)\!\bigg\}\qquad
\nonumber\\
\nonumber\\
c_3&=& \zeta[-4,\rho_2]\, (1/2\!-\!\rho_1)\!+\!\zeta[-4,\rho_1]\,
(1/2\!-\!\rho_2)\!+\!2 \zeta[-2,\rho_1]\,\zeta[-2,\rho_2],
\nonumber\\
\nonumber\\
c_2&=& (1/2\!-\!\rho_1)\,\zeta[-2,\rho_2]\!+
\!\zeta[-2,\rho_1]\,(1/2\!-\!\rho_2),
\qquad
c_1= (1/2\!-\!\rho_1) (1/2\!-\!\rho_2)\\[-10pt]
\nonumber
\end{eqnarray}
In the first  expression $\tilde\cD_f$ describes the terms
$\cO(\epsilon^0)$, and its exact value depends on whether $m L$ is
smaller or larger than unity, and is discussed later on. The zeta
functions appearing in the coefficients $c_i$ are given
by
\begin{eqnarray}
\zeta[-2,x]&= & -\frac{1}{6}\, x (x-1)(2 x-1)\nonumber\\
\zeta[-4,x]& = & -\frac{1}{30}\, x (x-1) (2x-1) (3 x^2-3x-1)
\end{eqnarray}
Writing
\begin{eqnarray}\label{totalv}
V_\cO= V_{\cO \infty}+ V^{\rm ren}_\cO
\end{eqnarray}
 we therefore identify the following UV divergent terms:

\begin{eqnarray}\label{vz4}
V_{\cO  \infty}= \frac{m^6\,\cA}{768 \pi^3 \,\epsilon} +
\frac{m^4\,\,\, c_1}{16 \pi^2\,\, \epsilon}
 + \frac{m^2\,\, c_2}{2
\cA\,\, \epsilon} + \frac{\pi^2\,\,\,c_3}{\cA^2\,\,\epsilon}
\\[-5pt]
\nonumber
\end{eqnarray}
Notice that the structure of these divergences is valid
independent of the relative size of $m$ and $\cA$. For the $Z_4$
orbifold  we have seen that consistency of the boundary conditions
requires we choose the value of $\rho_1 = \rho_2=\rho$ with $\rho$
equal to 0 or $\frac12$, and so we must evaluate the coefficients
$c_k$ with these choices. Since both $\zeta[-2,\rho]$ and
$\zeta[-4,\rho]$ vanish when $\rho = 0$ or $\rho = \frac12$, we
see that for all such cases
\begin{eqnarray}\label{cs}
c_2 = c_3 = 0
\end{eqnarray} leaving in $V_{\cO\,\infty}$  only the
divergences in the first and second terms in eq.(\ref{vz4}).

The first  term in (\ref{vz4}) is a renormalization of the bulk
cosmological constant and is present irrespective of the values of
$\rho_1$ or $\rho_2$. Its coefficient is $1/4$ the size of the
similar result for the covering torus. Therefore, once this term
is expressed in terms of the orbifold area, $\cA_{\cal O} =
\cA/4$, its coefficient is precisely the same as was found for
$\cT_2$ and $\cT_2/Z_2$, as expected. The term in (\ref{vz4}),
proportional to $c_1$, is a ``brane'' divergence, which
renormalizes the brane tension. It is nonzero only for the case
where $\rho_1 = \rho_2 = 0$, in which case $c_1 = \frac14$.

To conclude, we find that the UV divergences for the Casimir
energy due to compactifications on $\cT_2/Z_4$ with discrete
Wilson lines $(\rho_1,\rho_2)$, have two kinds of divergences at
one loop, similar to the case of  $\cT_2/Z_2$. These have the form
and coefficients required by the general heat-kernel analysis
\cite{Doug} and renormalize the bulk cosmological constant and the
tension of branes localized at the orbifold fixed points. For the
case of remaining orbifolds $\cT_2/Z_3$,  $\cT_2/Z_6$, the
analysis of the divergences of the quantity $\tilde\cD$ of
(\ref{ddefinition}) is more involved since the Zeta function
entering its definition will retain a $n_1$ dependence. This makes
the computation more tedious and the identification of the
relevant counterterms more difficult to analyze in this case
\cite{dmg}.

\vspace{0.8cm}
\subsubsection{The finite part of Casimir Energy.}

For the finite part of the Casimir energy  for
the orbifold  $\cT_2/Z_4$ one obtains in the limit of $m^2 \,\cA\ll 1$
(see Appendix eqs.(\ref{rrr}) and (\ref{qqq})).

\begin{eqnarray}\label{vorbz4}
V^{\rm ren}_\cO(\rho_1,\rho_2)& =& -\frac{1}{2 \,\cA^2}\,
\bigg\{\tilde\cD_f+
\frac{2 \pi^3}{45}  \bigg[\frac{1}{21}\!-\rho_1^2
  (1-5\rho_1^2 -\! 2 \rho_1^4+6 \rho_1^3)\bigg]\!
\\
\nonumber\\
&+&
\sum_{n_1<0}
\bigg[ (n_1+\rho_1)^2 \, \Li_3(\sigma_{n_1})
\!+\!\frac{3 \,\vert n_1+\rho_1\vert }{2\pi}\, \Li_4(\sigma_{n_1})
\!+\!\frac{3\,}{4\pi^2}  \, \Li_5(\sigma_{n_1})\!+\!c.c.
\bigg]\!\bigg\}\qquad\,\, \nonumber\\
\nonumber
\end{eqnarray}
with the notation
\begin{eqnarray}\label{sigmaz4}
\sigma_{n_1<0}= e^{2 \pi \big(n_1+\rho_1+i \rho_2\big)},\qquad
0\leq \rho_i<1
\end{eqnarray}
Here $\tilde\cD_f$ is an asymptotic series which for $m^2\cA\! \ll\! 1$
has the  following expression (see Appendix~\ref{appendixC}, eq.(\ref{rrr}))
\begin{eqnarray}\label{finitep}
\tilde\cD_f&=&\pi^2\bigg\{
2\zeta[-2,\rho_1]\zeta[-2,\rho_2]\ln(\tau \pi e^{\gamma-1})
+
\zeta[0,\rho_2]\Big(\zeta[-4,\rho_1]\ln(\pi\tau e^{\gamma-\frac{3}{2}})
+2  \zeta'[-4,1-\rho_1]\Big)
\nonumber\\
\nonumber\\
&+&
4 \zeta[-2,\rho_2] \zeta'[-2,1-\rho_1]
+\zeta[-4,\rho_2]
\Big(\zeta[0,\rho_1]\ln(\pi\tau e^\gamma)+2\zeta'[0,1-\rho_1]\Big)
\nonumber\\
\nonumber\\
&+& 2 \sum_{k\geq 3} \frac{(-1)^k}{k!}\Gamma[k-2] \zeta[-2 k,
\rho_2]\, \zeta[2 k-4, 1-\rho_1]\bigg\}, \quad {\rm with}\quad
\tau=(2\pi)^2/(\mu^2 \cA).
\end{eqnarray}
Since\footnote{In (\ref{finitep}) the derivative of Zeta function is
taken wrt its first argument.}
 for the orbifold $\cT_2/Z_4$ the Wilson  lines have the values
$\rho_1\!=\!\rho_2\!=\!0, 1/2$, $\tilde \cD_f$ simplifies to give:
\begin{eqnarray}\label{ss1}
\tilde\cD_f\Big\vert_{\rho_1=\rho_2=0}=\frac{3}{4\pi^2}\, \zeta[5],
\qquad {\rm and}\qquad
\tilde\cD_f\Big\vert_{\rho_1=\rho_2=\frac12}=0.
\end{eqnarray}
Eqs.(\ref{vorbz4}), (\ref{ss1}) and also (\ref{totalv}),
(\ref{vz4}), (\ref{cs}) give the final result for
 the  Casimir energy for the orbifold
$\cT_2/Z_4$ with discrete Wilson lines. Finally, with
 $\rho_1=\rho_2=0,1/2$ one has from eqs.(\ref{vorbz4}) to (\ref{ss1})
\begin{eqnarray}
V^{\rm ren}_\cO(\rho,\rho)=\frac{1}{4} \,V^{\rm ren}
(\rho,\rho)\vert_{m=0},\qquad \rho=0, \frac{1}{2}
\end{eqnarray}
where $V^{\rm ren}$ is the result for the 2-torus $\cT_2$, given  in eq.(\ref{wmd}). 
Following closely these steps, one can
also obtain from eqs.(\ref{vorb2}), (\ref{ddefinition}) similar results for
$\cT_2/Z_6$ and $\cT_2/Z_3$ orbifolds.

\subsubsection{Heavy-mass dependence.}
We discuss now the heavy mass dependence for the Casimir energy.
For $\cT_2/Z_4$  with  $m\! \gg \! 1/L$ it turns out that the
divergences in $V_\cO$ are identical to those in eq.(\ref{vz4}).
From Appendix \ref{appendixC}, eq.(\ref{erwsdfgv}) with
(\ref{lll}), (\ref{qqq}), (\ref{ptr}) one obtains the full result
for $V_\cO$ for $m L\!\gg \! 1$. Here we outline only the main
behaviour which is
\begin{eqnarray}\label{vlargeml}
V_{\cO}&=& - \bigg\{ \frac{m^4\, c_1}{32 \pi^2}\, \ln(\pi
e^{\gamma-\frac{3}{2}}m^2/\mu^2) + \frac{1}{2}\frac{m^6 \cA}{768
\pi^3}\ln(\pi  \, e^{\gamma-\frac{11}{6}} m^2/\mu^2)+\cdots\bigg\}
\end{eqnarray}
with $c_{1}=(1/2-\rho)^2$ and where the dots account for
additional terms such as polylogarithms terms, identical to those
in (\ref{vorbz4}), and for (asymptotic series of)  terms which are
suppressed by inverse powers of $m^2 \cA$.  The latter vanish in
the special case of $\cT_2/Z_4$ with $\rho=0, 1/2$, to leave only
the (exponentially suppressed) polylogarithm contributions.

\section{Conclusions}

In this paper we compute the value of the Casimir energy for a
very broad class of two-dimensional toroidal compactifications.
These include the general case of $\cT_2$ compactifications with
arbitrary boundary conditions for the 6D fields corresponding to
the presence of arbitrary Wilson lines, as well as $\cT_2/Z_N$
orbifolds (also with Wilson lines) obtained by identifying points
under $Z_N$ rotations. Our calculations are explicit for a 6D
scalar having an arbitrary 6D mass $m$, and we show how to extend
these results to higher-spin fields for supersymmetric 6D
theories. Particular attention was paid to regularization issues
and to the identification of the divergences of the potential.
The computation also investigated the dependence of the result on
$m$, including limits for which $m^2 \cA$ is larger or smaller
than unity, (where $\cA$ is the volume of the internal 2
dimensions).

For the cases of $\cT_2$ and $\cT_2/Z_2$, our calculation
generalizes earlier results to include the dependence on an
arbitrary complex structure, $U$, for the underlying torus. The
potential obtained is likely to be useful for studies of the
dynamics of these moduli, including their stabilization and their
potential applications to cosmology \cite{Ponton:2001hq,ABRS2}.

By carefully isolating the UV divergent part of $V$, we show that
all of the divergences may be renormalized into a bulk
cosmological constant (which gives a Casimir energy proportional
to $m^6 \cA$) and - for the case of $\cT_2/Z_2$ - a cosmological
constant (or brane tension) localized at the orbifold fixed points
(which gives a Casimir energy proportional to $m^4$). Furthermore,
these divergences agree with expectations based on general
heat-kernel calculations, such as those recently performed for 6D
compactifications in ref.~\cite{Doug}. For massive 6D scalar
fields, $m^2 \cA\gg 1$, the dependence on $m$ of the finite part
of the Casimir energy obtained in the modified minimal
subtraction scheme,  is exponentially suppressed.

We present results for the Casimir energy also for $\cT_2/Z_N$
orbifolds with $N>2$, again including Wilson lines and any shape
moduli which are allowed. The case $\cT_2/Z_4$ was studied in
particular detail. The UV divergences that emerge in this case
again take the form required by general heat-kernel arguments, and
can be absorbed into renormalizations of the bulk cosmological
constant and brane tensions localized at the orbifold fixed
points. The finite part of the Casimir energy was computed in
detail and may be used for phenomenological applications. Finally,
the technical tools  of the Appendix can be used for other
applications such as the one-loop corrections to the  gauge
couplings in gauge theories on orbifolds, in the presence of
discrete Wilson lines.

\vspace{0.6cm} \noindent {\bf Acknowledgements.}\newline We thank
Y. Aghababaie, Z. Chacko, J. Elliot, G. Gabadadze and A. de la
Macorra for helpful discussions about 6D Casimir energies on the
torus. D.G. thanks S.~Groot-Nibbelink and Hyun~Min~Lee for 
discussions on some technical aspects of this work.

\vspace{0.3cm} \noindent C.B.'s research is supported by grants
from NSERC (Canada) and McMaster University and D.H. acknowledges
partial support from McGill University. The research of F.Q. is
partially supported by PPARC and a Royal Society Wolfson award.
The work of D.~Ghilencea was supported by a post-doctoral research
grant from Particle Physics and Astronomy Research Council
(PPARC), United Kingdom. D.~Ghilencea acknowledges the support
from the RTN European program MRTN-CT-2004-503369,  
to attend the ``Planck 2005'' conference 
where this work was completed.

\newpage  
\section*{Appendix}
\def\theequation{\thesubsection-\arabic{equation}}
\setcounter{equation}{0}
\def\thesubsection{A}

\subsection{.\,\, Calculation of the
vacuum energy in DR for 2D compactifications.} \label{appendixA}
\setcounter{equation}{0} We provide here details of the
calculation of the vacuum energy. One has ($d=4-\epsilon$)

\begin{eqnarray}\label{eq1}
\!\!\! V^*(\rho_1,\rho_2) \!\equiv\! \mu^{4-d}\! \!\!
\sum_{n_{1,2}\in\bZ}' \int\! \frac{d^dp}{(2\pi)^d} \ln\Big[ p^2 +
M_{n_1,n_2}^2 \Big]\!
=\frac{-\mu^{4}}{(2\pi)^{d}}\!\!\!\sum_{n_{1,2}\in\bZ}'
\int_0^\infty \!\!\!\!\! \frac{dt}{t^{1+d/2}} e^{-\pi\,t\,
\big[M_{n_1,n_2}^2/\mu^2\big]}\,\,\,\,
\\
\nonumber
\end{eqnarray}
 $\mu$ is a {\it finite, non-zero} mass scale introduced by the
DR scheme.  A ``prime'' on a double sum excludes the
$(n_1,n_2)=(0,0)$ mode. If a level $(n_1,n_2)$ is massless
$M_{n_1,n_2}=0$ (for example if $\rho_{1,2}\in \bZ$)),
mathematical consistency requires one shift
$M_{n_1,n_2}^2\rightarrow M_{n_1,n_2}^2+m^2$ by a finite non-zero
$m^2=\delta\,\mu^2$ ($\delta$ dimensionless). This also helps us
identify the scale ($m$) dependence of the divergences (poles in
$\epsilon$). We  use
\begin{equation}\label{mass2}
M^2_{n_1,n_2}=
\frac{(2\pi)^2}{\cA \,U_2} \vert
n_2+\rho_{2}-U(n_1+\rho_{1})\vert^2; \quad U\equiv U_1+i U_2 =
e^{i\theta} \frac{L_2}{L_1};\quad
\cA=L_1 L_2 \sin\theta
\end{equation}
The DR regularized sum in (\ref{eq1}) is re-written
\begin{eqnarray}\label{vj}
\!\!\!\!\! V^*(\rho_1,\rho_2) = \mu^4 \, C_\epsilon\!\!
\sum_{n_{1,2}\in\bZ}' \int_0^\infty\!\!\frac{dt}{t^{3-\epsilon/2}}
\, e^{-\pi \, t\, [M_{n_1,n_2}^2/\mu^2+ \delta ]} \equiv  \mu^4 \,
C_\epsilon\,\cJ^*_{-\epsilon/2},\quad
C_\epsilon=\frac{-1}{(2\pi)^{4-{\epsilon}}}\,\,
\end{eqnarray}
with $\delta=m^2/\mu^2$. The calculation of $V^*$ is reduced to
 that of  $\cJ_\epsilon^*$ performed  below.

\vspace{1cm}
\def\thesubsection{B}
\setcounter{equation}{0}
\subsection{.\,\, Series of Kaluza-Klein integrals and their DR regularization.}
\label{appendix2} We  evaluate (in the text $\tau \ra
(2\pi)^2/(\mu^2 \cA\, U_2)$,
$\delta\ra m^2/\mu^2$)

\begin{eqnarray}
\!\!\! \!\!\!\cJ_\epsilon^*\equiv\!\!\!\!
 \sum_{n_1,n_2\in\bZ}' \int_{0}^{\infty}\!\!\! \frac{dt}{t^{3+\epsilon}} \,\,
e^{-\pi t\, \tau  \vert n_2+\rho_{2}- U (n_1+\rho_{1})\vert^2}\,
e^{-\pi\, \delta\, t},\,\,\, \tau\!>\! 0,\,\, U\!\equiv\! U_1\!
+\! i U_2,\,\, \delta\!>\!0,\,\,
\rho_i\!\in \!\bR\,\,\label{ldef}\\
\nonumber
\end{eqnarray}

\noindent $\cJ_\epsilon^*$ includes shape moduli effects
($\theta\!\not=\!\pi/2$, $L_1\!\not=\! L_2$), $\delta$ shifts, and
arbitrary ``twists'' $\rho_i$ wrt $L_{1,2}$. To evaluate
$\cJ_\epsilon^*$ one uses  re-summation (\ref{p_resumation}); the
integrand becomes

\begin{eqnarray}\label{sums}
\!\!\sum_{n_1,n_2\in\bZ}'e^{-\pi t \, \tau \vert n_2+\rho_{2}- U
(n_1+\rho_{1})\vert^2 } \!&=&\!\sum_{n_2\in\bZ}'e^{-{\pi\, t
\,\tau}  |n_2+\rho_2-U\rho_1|^2}
+\sum_{n_1\in\bZ}'\sum_{n_2\in\bZ} e^{-{\pi\, t\, \tau
|n_2+\rho_2-U(n_1+\rho_1)|^2}}
\nonumber\\
\nonumber\\
\!&=&\!
  \sum_{n_2\in\bZ}' e^{- \pi\, t\, \tau  |n_2+\rho_2-U \rho_1|^2}+
  \frac{1}{\sqrt{t\,\tau}} \sum_{n_1\in\bZ}'
e^{-\pi t \, \tau U_2^2  \, (n_1+\rho_1)^2}
\nonumber\\
\nonumber\\
+&&\!\!\!\!\!\!\frac{1}{\sqrt{t\,\tau}}\!\!
\sum_{n_1\in\bZ}'\sum_{\tilde n_2\in\bZ}' e^{-\frac{\pi {\tilde
n_2}^2}{t\,\tau} -\pi t\,\tau  U_2^2 \, (n_1+\rho_1)^2+2 \pi i
{\tilde n_2} (\rho_2-U_1(\rho_1+n_1))}
\\
\nonumber
\end{eqnarray}
A prime on a double sum indicates  that $(n_1,n_2)\!\not=\!(0,0)$
is excluded and a ``prime'' on a single sum excludes its $n=0$
mode. The three contributions above can be integrated termwise for
any real $\rho_{i}$, (given the presence of $e^{-\pi\, t\,
\delta}$).  Accordingly, one has  three contributions
\begin{eqnarray}\label{j1j2j3}
\cJ_\epsilon^*(\rho_1,\rho_2) &=&
\cK_1(\rho_1,\rho_2)+\cK_2(\rho_1,\rho_2)+\cK_3(\rho_1,\rho_2)
\end{eqnarray}
defined/evaluated in the following:

\vspace{0.2cm} \noindent $\bullet$ Computing $\cK_1$:
\begin{eqnarray}\label{kk1}
\cK_1&\equiv &\sum_{n_2\in\bZ}'
\int_{0}^{\infty}\frac{dt}{t^{3+\epsilon}}
 e^{- \pi\, t\,\tau  \vert n_2+\rho_2-U\rho_1\vert^2}\,
e^{-\pi\, \delta\, t} =
\frac{\pi^2}{2 \,\epsilon} \,(\delta+\tau\vert \rho_2-U \rho_1\vert^2)^2\\
\nonumber\\
&+&\frac{\pi^2}{4}(\delta+\tau \vert \rho_2 -U \rho_1\vert^2)^2
\ln\bigg[\pi^2 \, e^{2\gamma-3}\, (\delta+\tau \vert \rho_2 -U
\rho_1\vert^2)^2\bigg] - \frac{8\pi^3}{15\sqrt \tau} (\delta+\tau
U_2^2 \rho_1^2)^{\frac{5}{2}}
\nonumber\\
\nonumber\\
+&&\!\!\! \!\!\!\!\! \tau (\delta+\tau \, U_2^2 \rho_1^2)\,
\Li_3(e^{-2\pi\gamma(0)}) +\frac{3 \,\tau ^{\frac{3}{2}}}{2\pi}\,
(\delta+\tau \, U_2^2 \rho_1^2)^{\frac{1}{2}}
\,\Li_4(e^{-2\pi\gamma(0)}) +\frac{3\, \tau^2}{4\pi^2}\,
\Li_5(e^{-2\pi\gamma(0)}) \! +\! c.c.
\nonumber\\
\nonumber
\end{eqnarray}
where ``c.c.'' applies to the PolyLogarithm functions only. To
evaluate $\cK_1$ we first added and subtracted the $n_2=0$ mode
 contribution.
We then  used a  (Poisson) re-summation over $n_2$, then the
integral representation of modified Bessel functions
(\ref{bessel1}) with (\ref{k52}),  and  the definition of the
 Polylogarithm $\Li_\sigma(x)$ (\ref{plg}). Finally we used the notation
\begin{eqnarray}
\gamma(0)\equiv \frac{1}{\sqrt\tau}(\delta+\tau U_2^2 \rho_1^2
)^{\frac{1}{2}} -i (\rho_2-U_1 \rho_1)
\end{eqnarray}
The divergence of $\cK_1$ is that of the  {\it  excluded} $n_2=0$
mode in  $\cK_1$, which is in turn due to the absence of
$(n_1,n_2)=(0,0)$ in the definition of $\cJ_\epsilon^*$.

\vspace{0.3cm} \noindent $\bullet$ Computing $\cK_2$: \,\,\, We
introduce the notation $\Delta_{\rho_1}\!\equiv\!
\rho_1\!-\![\rho_1]$, $0\!\leq\! \Delta_{\rho_1}\!<\! 1$,
$[\rho_1]\!\in\!\bZ$.

\noindent {\bf (a). } For $0\!\leq \!\delta/(\tau U_2^2)\!<\! 1$
we have
\begin{eqnarray}
\cK_2\! &\equiv & \!\!\frac{1}{\sqrt{\tau}}\sum_{n_1\in\bZ}'
\int_{0}^{\infty}\!\frac{dt}{t^{7/2+\epsilon}} \,e^{-\pi t\,\tau
\, U_2^2\,  (n_1+\rho_1)^2}\, e^{ -\pi \, \delta \, t}
\nonumber\\
\nonumber\\
=&&\!\!\!\! \frac{1}{\sqrt{\tau}} \int_0^\infty
\frac{dt}{t^{7/2+\epsilon}} \bigg[\sum_{n_1\in\bZ} e^{-\pi\, t\,
[\delta+\tau U_2^2
    (n_1+\Delta_{\rho_1})^2]}
-e^{-\pi\, t\, (\delta+\tau U_2^2 \rho_1^2)}\bigg]
\nonumber\\
\nonumber\\
=&&\!\!\!\!\!\! \frac{\pi^{5/2+\epsilon}}{\sqrt{\tau}} \,
\Gamma[\!-5/2-\!\epsilon] \bigg[ \sum_{n_1\in\bZ}' [\delta+\tau
U_2^2 (n_1+\Delta_{\rho_1})^2]^{\frac{5}{2}+\epsilon} \! - \!
(\delta+\tau U_2^2 \rho_1^2)^{\frac{5}{2}+\epsilon} \! +\!
(\delta+\tau U_2^2 \Delta_{\rho_1}^2)^{\frac{5}{2}+\epsilon}\bigg]
\nonumber\\
\nonumber\\
=&&\!\!\!\! \frac{\pi^{\frac{5}{2}}}{\sqrt\tau}\, \Gamma[-5/2]
\bigg[(\delta+\tau U_2^2 \Delta_{\rho_1}^2)^{\frac{5}{2}}
-(\delta+\tau U_2^2 {\rho_1}^2)^{\frac{5}{2}}\bigg]
\nonumber\\
\nonumber\\
+&&\!\!\!\! \!\!\!\frac{(\pi\, \tau
U_2^2)^{\frac{5}{2}+\epsilon}}{\sqrt\tau} \sum_{k\geq 0}
\frac{\Gamma[k-5/2-\epsilon]}{k!} \bigg[\frac{-\delta}{\tau
U_2^2}\bigg]^k \bigg[\zeta[2 k\!
-5-\!2\epsilon,1+\!\Delta_{\rho_1}]+ (\Delta_{\rho_1}\rightarrow
-\!\Delta_{\rho_1})\bigg] \label{k2exp}
\end{eqnarray}
In the last step  we  used the  binomial
 expansion
\begin{equation}\label{zp}
\sum_{n\geq 0}[a(n+c)^2+q]^{-s}=a^{-s}\sum_{k\geq 0}
\frac{\Gamma[k+s]}{k\,!\, \Gamma[s]} \bigg[\frac{-q}{a}\bigg]^k
\zeta[2 k+ 2 s, c],\qquad 0<q/a\leq 1
\end{equation}
Here  $\zeta[q,a]$ with  $a\not=0,-1,-2,\cdots$ is the Hurwitz
zeta function, (with $\zeta[q,a]=\sum_{n\geq 0} (a+n)^{-q}$ for
$\textrm{Re}(q)>1$). Hurwitz zeta-function has one singularity
(simple pole) at $q=1$ and  $\zeta[q,1]=\zeta[q]$ with $\zeta[q]$
the Riemann zeta function. The  only divergence in $\cK_2$ is due
to its  $k=3$ term in (\ref{k2exp}), from  the singularity of the
Zeta function. In the remaining terms in the series one can safely
set $\epsilon=0$. Further
\begin{eqnarray}\label{expansionzeta}
\zeta[1-2\epsilon,1\pm \Delta_{\rho_1}]& = & -\frac{1}{2\epsilon}
-\psi(1\pm \Delta_{\rho_1})+\cO(\epsilon)\nonumber\\
\Gamma[1/2-\epsilon] &=& \pi^{1/2}(1+\epsilon \ln(4
e^\gamma))+\cO(\epsilon)
\nonumber\\
x^\epsilon &= & 1+\epsilon \ln x+\cO(\epsilon)
\end{eqnarray}
we find for $0\leq \delta/(\tau U_2^2)<1$:
\begin{eqnarray}\label{kk2}
\cK_2 & =& \frac{\pi^3 \delta^3}{6\tau \vert U_2\vert}
\frac{1}{\epsilon}+ \frac{\pi^3 \delta^3}{6\tau \vert U_2\vert}
\ln\bigg[4\pi\tau U_2^2 e^{\gamma+\psi(\Delta_{\rho_1})+
\psi(-\Delta_{\rho_1})}\bigg]
\nonumber\\
\nonumber\\
-&&\!\!\!\! \!\frac{8\pi^3}{15\sqrt\tau} \bigg[(\delta+\tau U_2^2
\Delta_{\rho_1}^2)^{\frac{5}{2}}- (\delta+\tau U_2^2
{\rho_1}^2)^{\frac{5}{2}}+(\tau U_2^2)^{\frac{5}{2}}
(\zeta[-5,1\!+\Delta_{\rho_1}]+
\zeta[-5,1\!-\Delta_{\rho_1}])\bigg]\nonumber\\
\nonumber\\
+&&\!\!\!\! \frac{-4 \pi^3}{3\sqrt\tau} \,\delta\, \tau^{3/2} \,
\vert U_2\vert^3 (\zeta[-3,1+\Delta_{\rho_{1}}]+
\zeta[-3,1-\Delta_{\rho_{1}}]) +\pi^3 \, \delta^2\,
 \vert U_2\vert \, (1/6+\Delta_{\rho_1}^2)
\nonumber\\
\nonumber\\
+&&\!\!\!\! \frac{\pi^{5/2}}{\sqrt\tau} (\tau U_2^2)^{5/2}
\sum_{p\geq 1} \frac{\Gamma[p+1/2]}{(p+3)!}
\bigg[\frac{-\delta}{\tau U_2^2}\bigg]^{p+3}
\bigg[\zeta[2p+1,1+\Delta_{\rho_1}]+\zeta[2p+1,1-\Delta_{\rho_1}]\bigg]
\\
\nonumber
\end{eqnarray}
The  divergence in $\cK_2$ is due to $\tilde n_2=0$ (i.e. Poisson
re-summed zero mode wrt to the second dimension) in the presence
of infinitely many KK modes of the first dimension ($n_1$). It is
thus an interplay effect of both compact dimensions. The condition
of validity of the above result $0\leq \delta/(\tau U_2^2)<1$
gives $\delta \,\mu^2\leq 1/L_{1,2}^2$. If $\delta\!\ll\! 1$ the
result (\ref{kk2}) simplifies considerably.

\noindent {\bf (b).} If $\delta/(\tau U_2^2)\!\geq \! 1$ or
$\delta\!\gg\! 1$ eq.(\ref{zp}) does not converge. If so,  $\cK_2$
is reevaluated as below:
\begin{eqnarray}
\!\!\!\!\!\cK_2\! &\equiv &
\!\!\frac{1}{\sqrt{\tau}}\sum_{n_1\in\bZ}'
\int_{0}^{\infty}\!\frac{dt}{t^{7/2+\epsilon}} \,e^{-\pi t\,\tau
\, U_2^2\,  (n_1+\rho_1)^2}\, e^{ -\pi \, \delta \, t}
\nonumber\\
\nonumber\\
=&&\!\!\!\!\! \frac{1}{\sqrt{\tau}} \int_0^\infty
\frac{dt}{t^{7/2+\epsilon}} \bigg[\sum_{n_1\in\bZ} e^{-\pi\, t\,
[\delta+\tau U_2^2
    (n_1+{\rho_1})^2]}
-e^{-\pi\, t\, (\delta+\tau U_2^2 \rho_1^2)}\bigg]
\nonumber\\
\nonumber\\
=&& \!\!\!\!\!\!\!\! \frac{1}{\sqrt\tau}\! \int_0^\infty\!\!\!\!\!
\frac{dt}{t^{7/2+\epsilon}}\bigg[\frac{1}{\sqrt{t \tau}
      U_2 }\sum_{\tilde n_1\in \bZ}' e^{-\frac{\pi \tilde n_1^2}{t
      \tau U_2^2}+2 i \pi \tilde n_1 \rho_1 -\pi \delta t}\!\!\!- \! e^{-\pi t
    (\delta+\tau U_2^2 \rho_1^2)}\bigg]\!\! +\! \!
\frac{(\pi \delta)^{3+\epsilon}}{\tau U_2}
\Gamma[-3-\epsilon]\qquad
\\
\nonumber
\end{eqnarray}
where the last term originates in the   $\tilde n_1=0$ term of the
series. We find (with (\ref{bessel1}))
\begin{eqnarray}\label{diff}
\cK_2 & = & \frac{\pi^3 \delta^3}{6 \tau \vert U_2 \vert}
\Big[\frac{1}{\epsilon}+\ln\Big(\pi \delta e^{\gamma
-11/6}\Big)\Big] + 4 \sqrt\tau
U_2^2\delta^{\frac{3}{2}}\sum_{\tilde n_1\geq 1}\frac{\cos[2\pi
  \tilde n_1\rho_1]}{\tilde n_1^3}\, K_3 \bigg(\! 2\pi \tilde n_1
\sqrt{\delta/(\tau U_2^2)}\!\bigg)\,\,\,\,
\nonumber\\
\nonumber\\
&+& \frac{8 \pi^3}{15\sqrt\tau}(\delta+\tau U_2^2
\rho_1^2)^{\frac{5}{2}}
\end{eqnarray}
 rapidly convergent
 if $\delta\!\geq\! \tau U_2^2$ or  $\delta\!\gg\! 1$.
 This ends our calculation of $\cK_2$ at large/small~$\delta$.

\vspace{0.3cm} \noindent $\bullet$ Computing $\cK_3$:
\begin{eqnarray}\label{kkk3}
\cK_3 \equiv \frac{1}{\sqrt{\tau}}\sum_{n_1\in\bZ}'\sum_{\tilde
n_2\in\bZ}' \int_{0}^{\infty}\!\frac{dt}{t^{7/2+\epsilon}} \,
e^{-\pi \tilde n_2^2/(t\,\tau) -\pi t\,\tau\,  U_2^2
\,(n_1+\rho_1)^2+2 \pi i \tilde n_2 (\rho_2-U_1
(\rho_1+n_1)) -\pi \delta\, t}\\
\nonumber
\end{eqnarray}
Since $\cK_3$ is always exponentially suppressed at $t\rightarrow
0$ and at $t\rightarrow \infty$ it has no singularities. We can
thus safely set $\epsilon=0$. One finds
\begin{eqnarray}\label{kk3}
\cK_3\! &\! = \!&\! {\tau}\!\! \sum_{n_1\in\bZ}'\! \bigg[ z(n_1)
\, \Li_3(e^{-2\pi
  \gamma(n_1)})
\!+\!\frac{3}{2\pi} (\tau z(n_1))^{1/2}\, \Li_4(e^{-2\pi \,
\gamma(n_1)}) +\!\frac{3 \tau }{4\pi^2}  \,
\Li_5(e^{-2\pi\gamma(n_1)}) \bigg]\!+c.c. \nonumber
\end{eqnarray}
\begin{eqnarray}\label{zz}
z(n_1)& \equiv & \delta+\tau U_2^2 (n_1+\rho_1)^2
\nonumber\\
\nonumber\\
\gamma(n_1) & \equiv & \frac{1}{\sqrt\tau}(\delta+\tau U_2^2
(n_1+\rho_1)^2)^{1/2}- i(\rho_2-U_1(n_1+\rho_1))
\end{eqnarray}
To evaluate $\cK_3$ we used the  representation of Bessel
functions eq.(\ref{bessel1}), then (\ref{k52}) and finally
 the polylogarithm definition  in (\ref{plg}).
This result simplifies considerably if $\delta\ll 1$.

\vspace{0.3cm} To conclude if $0\!\leq\! \delta/(\tau
U_2^2)\!<\!1$ we find for  $\cJ_\epsilon^*$ (with (\ref{ldef}),
(\ref{j1j2j3}), (\ref{kk1}), (\ref{kk2}), (\ref{zz}))

\begin{eqnarray}\label{a16}
\cJ_\epsilon^* &\equiv &
 \sum_{n_1,n_2\in\bZ}' \int_{0}^{\infty} \frac{dt}{t^{3+\epsilon}} \,
e^{-\pi t\, \tau  \vert n_2+\rho_{2}- U (n_1+\rho_{1})\vert^2}\,
e^{-\pi\, \delta\, t}
\nonumber\\
\nonumber\\
= &&\!\!\!\! \frac{\pi^2}{2 \,\epsilon} \,(\delta+\tau\vert
\rho_2-U \rho_1\vert^2)^2 +\frac{\pi^2}{4}(\delta+\tau \vert
\rho_2 -U \rho_1\vert^2)^2 \ln\bigg[\pi^2 \, e^{2\gamma-3}\,
(\delta+\tau \vert \rho_2 -U \rho_1\vert^2)^2\bigg]
\nonumber\\
\nonumber\\
+&&\!\!\!\!\! {\tau}\!\! \sum_{n_1\in\bZ} \bigg[ z(n_1) \,
\Li_3(e^{-2\pi
  \gamma(n_1)})
+\frac{3}{2\pi} (\tau z(n_1))^{1/2}\, \Li_4(e^{-2\pi \,
\gamma(n_1)}) +\frac{3 \tau }{4\pi^2}  \,
\Li_5(e^{-2\pi\gamma(n_1)}) \bigg]\!+\! c.c.
\nonumber\\
\nonumber\\
- &&\!\!\!\!\! \frac{8\pi^3}{15\sqrt \tau}\, (\delta\!+\!\tau
U_2^2\Delta_{\rho_1}^2)^{\frac{5}{2}} + \!\frac{\pi^3
\delta^3}{6\tau \vert U_2\vert}
\bigg[\frac{1}{\epsilon}\!+\!\ln\Big[4\pi\tau U_2^2
e^{\gamma+\psi(\Delta_{\rho_1})+\psi(-\Delta_{\rho_1})}
\Big]\bigg] \!+\!\pi^3 \delta^2 \vert U_2\vert
\big(\frac{1}{6}+\Delta_{\rho_1}^2\big)
\nonumber\\
\nonumber\\
+&&\!\!\!\! \frac{4 \pi^3}{45} \tau^2\, \vert U_2\vert^5 \Big[
\frac{1}{21}-\Delta_{\rho_1}^2
\,(1-5\Delta_{\rho_1}^2-2\Delta_{\rho_1}^4) \Big]- \frac{ \pi^3
\delta\, \tau \vert U_2 \vert^3}{45} \Big[ 1-30\,
\Delta_{\rho_1}^2 (1+\Delta_{\rho_1}^2)\Big]
\nonumber\\
\nonumber\\
+&&\!\!\!\! \pi^{5/2} \, \tau^2 \vert U_2\vert^5 \sum_{p\geq 1}
\frac{\Gamma[p+1/2]}{(p+3)!} \bigg[\frac{-\delta}{\tau
U_2^2}\bigg]^{p+3}
\bigg[\zeta[2p+1,1+\Delta_{\rho_1}]+\zeta[2p+1,1-\Delta_{\rho_1}]\bigg]
\end{eqnarray}
This restriction  $0\leq \delta/(\tau U_2^2)<1$ is required for
the convergence of the calculation of $\cK_2$. The first line in
$\cJ_\epsilon^*$ is due to the absence  of the mode $(0,0)$.
$\cJ_\epsilon^*$ is well defined even for $\delta=0$ if
$M_{n_1,n_2}\not=0$. In such case  the result is obtained by
redoing the above calculation with $\delta=0$ or more  easily,
from the one above by formally setting $\delta=0$. The above
result simplifies considerably when  $\delta\ll 1$.

Finally, if we have $\delta/(\tau U_2^2)\geq 1$, from
eqs.(\ref{ldef}), (\ref{j1j2j3}), (\ref{kk1}), (\ref{diff}),
(\ref{zz}) we find

\begin{eqnarray}\label{a17}
\cJ_\epsilon^* &\equiv &
 \sum_{n_1,n_2\in\bZ}' \int_{0}^{\infty} \frac{dt}{t^{3+\epsilon}} \,
e^{-\pi t\, \tau  \vert n_2+\rho_{2}- U (n_1+\rho_{1})\vert^2}\,
e^{-\pi\, \delta\, t}
\nonumber\\
\nonumber\\
 = &&\!\!\!\!
\frac{\pi^2}{2 \,\epsilon} \,(\delta+\tau\vert \rho_2-U
\rho_1\vert^2)^2 +\frac{\pi^2}{4}(\delta+\tau \vert \rho_2 -U
\rho_1\vert^2)^2 \ln\bigg[\pi^2 \, e^{2\gamma-3}\, (\delta+\tau
\vert \rho_2 -U \rho_1\vert^2)^2\bigg]
\nonumber\\
\nonumber\\
+&&\!\!\!\!\! {\tau}\!\! \sum_{n_1\in\bZ} \bigg[ z(n_1) \,
\Li_3(e^{-2\pi
  \gamma(n_1)})
+\frac{3}{2\pi} (\tau z(n_1))^{1/2}\, \Li_4(e^{-2\pi \,
\gamma(n_1)}) +\frac{3 \tau }{4\pi^2}  \,
\Li_5(e^{-2\pi\gamma(n_1)}) \bigg]\!+\! c.c.
\nonumber\\
\nonumber\\
+&&\!\!\!\!\! \frac{\pi^3 \delta^3}{6 \tau \vert U_2 \vert}
\Big[\frac{1}{\epsilon}+\ln\Big(\pi \delta e^{\gamma
-11/6}\Big)\Big] + 4 \sqrt\tau U_2^2\delta^{\frac{3}{2}}
\!\!\sum_{\tilde n_1\geq 1}\frac{\cos[2\pi
    \tilde n_1\rho_1]}{\tilde n_1^3}\, K_3 \bigg(2\pi \tilde n_1
\sqrt{\delta/(\tau U_2^2)}\bigg)
\end{eqnarray}
and this concludes the evaluation of $\cJ_\epsilon^*$. Using
(\ref{k52}), one shows that the last equation has the
 contributions from $K_3$ and from  the polylogarithms suppressed
if $\delta\gg \tau U_2^2$ and $\delta\gg \tau$, to leave the first
line and the term ${\pi^3 \delta^3}/({6 \epsilon \tau \vert U_2
\vert})$ as its leading behaviour for this region of the parameter
space.

\vspace{0.5cm} \noindent $\bullet$  Adding to $\cJ_\epsilon^*$ the
effect of
 the mode $(0,0)$ the result is
\begin{eqnarray}\label{jps}
\cJ_\epsilon(\rho_1,\rho_2)\equiv \cJ_\epsilon^*(\rho_1,\rho_2)
 + \cZ_\epsilon(\rho_1,\rho_2)\,\,
\end{eqnarray}
with
\begin{eqnarray}\label{zmode}
\cZ_\epsilon(\rho_1,\rho_2) & = &
 \int_{0}^{\infty} \frac{dt}{t^{3+\epsilon}} \,
e^{-\pi t\, \tau  \vert \rho_{2}- U \rho_{1})\vert^2}\,
e^{-\pi\, \delta\, t}\\
\nonumber\\
=&&\!\!\!\!\!\!\!\! \frac{-\pi^2}{2 \,\epsilon}(\delta+\tau\vert
\rho_2-U \rho_1\vert^2)^2\! -\frac{\pi^2}{4}(\delta+\tau \vert
\rho_2 -U \rho_1\vert^2)^2 \ln\bigg[\pi^2 \, e^{2\gamma-3}
(\delta+\tau \vert \rho_2 \!-\! U \rho_1\vert^2)^2\bigg]
\nonumber\\
\nonumber
\end{eqnarray}
$\cJ_\epsilon$ is thus  given by $\cJ^*_\epsilon$ without the
first  line in (\ref{a16}) or (\ref{a17}).

\vspace{0.3cm} \noindent
 Eqs. (\ref{uvdivergent}), (\ref{wmd}) in the text are then obtained from
\begin{eqnarray}\label{tz}
V(\rho_1,\rho_2) =  \mu^4 \,C_\epsilon
\,\cJ_{-\epsilon/2}(\rho_1,\rho_2),\quad\rm{with}\quad \delta\ra
m^2/\mu^2,\qquad  \tau \ra (2\pi)^2/(\mu^2 \cA U_2).
\end{eqnarray}
for the case $\delta\ll 1$.

\newpage
\def\theequation{\thesubsection-\arabic{equation}}
\def\thesubsection{C}
\setcounter{equation}{0}
\subsection{.\,\, More series of Kaluza-Klein  integrals
for $T_2/Z_N$ orbifolds.} \label{appendixC}

For general orbifolds $\cT_2/Z_N$ one needs to evaluate ($U\equiv
U_1+i\, U_2$)

\begin{eqnarray}\label{sss}
\cL_s& =& \sum_{n_1<0, \, n_2\geq 0}\int_0^\infty
\frac{dt}{t^{1-s}} \, e^{-\pi t \tau \vert n_2+\rho_2 -U
(n_1+\rho_1)\vert^2} \, e^{-\pi
  t \delta}
\nonumber\\
\nonumber\\
&=&\sum_{n_1<0,\, n_2\geq 0} \Gamma[s]\, \pi^{-s} \Big( \tau \vert
n_2+\rho_2 -U (n_1+\rho_1)\vert^2+\delta\Big)^{-s}\!\!,\quad \tau,
\delta>0,\,\, 0\!\leq\! \rho_{1,2}\!<\!1.
\\
\nonumber
\end{eqnarray}
Eq.(\ref{vorb2}) in the text  is then
$V_\cO=-\mu^4/(2\pi)^{4-\epsilon} \cL_{-(2-\frac{\epsilon}{2})}
\big(\delta\!\ra\! {m^2}/{\mu^2},\tau\!\ra\! (2\pi)^2/(\mu^2\cA
U_2)\big)$.

To compute $\cL_s$, the  usual (Poisson) re-summation used in
previous sections is not applicable given the restricted summation
on $n_1, n_2$. The sum of the  last line is actually a
``truncated'' Epstein function. To analyze $\cL_s$,  we follow the
method  in \cite{elizalde2},  for both non-zero $\rho_{1,2}$ and
complex $U$. For $s=-(2+\epsilon)$
 this allows us to evaluate the Epstein
function up to order $\cO(\epsilon^2)$, giving an expression for
$\cL_{-(2+\epsilon)}$ up to $\cO(\epsilon)$. We use that

\begin{eqnarray}\label{sume1}
E_1[z;\, s;\, \tau,\, c_1] & \equiv &  \sum_{n_2\geq 0}
[z\,+\,\tau  (n_2+c_1)^2]^{-s}
\end{eqnarray}
has the  asymptotic  expansion \cite{elizalde}
\begin{eqnarray}\label{ff1}
E_1[z; s; \tau, c_1] & \approx & {z^{-s}}\sum_{m\geq 0}
\frac{\Gamma[s+m]}{m!\, \Gamma[s]} \bigg[\frac{-\tau}{z}\bigg]^m
\zeta[-2 m, c_1]
+\frac{z^{1/2-s}}{2}\bigg[\frac{\pi}{\tau}\bigg]^{\frac{1}{2}}
\frac{\Gamma[s-1/2]}{\Gamma[s]}
\nonumber\\
\nonumber\\
& + & \frac{2 \pi^s}{\Gamma[s]} \tau^{-\frac{s}{2}-\frac{1}{4}}
z^{-\frac{s}{2}+\frac{1}{4}} \sum_{p\geq 1} p^{s-\frac{1}{2}}
\cos(2 \pi c_1 p) \, K_{s-\frac{1}{2}}\bigg(2\pi p
\big({z}/{\tau}\big)^{\frac{1}{2}}\bigg)\\
\nonumber
\end{eqnarray}
Here $\zeta[q,a], a\not=0,-1,-2, \cdots$ is the Hurwitz Zeta
function, $K_s$ is the Bessel function (\ref{bessel1}). In the
$m=0$ term in (\ref{ff1}) one can use $\zeta[0,c_1]=1/2-c_1$ even
for $c_1=0$. One can use (\ref{ff1}) recurrently for the 2D case
\cite{elizalde2}. With the substitutions
\begin{eqnarray}
c_1\ra c_1(n_1)\equiv\rho_2- U_1 (n_1+\rho_1);\, \qquad z\ra
z(n_1) \equiv \tau U_2^2 (n_1+\rho_1)^2+\delta
\end{eqnarray}
in eq.(\ref{ff1}) and  after applying a summation over $n_1$, one obtains
from (\ref{ff1}) $\cL_s$ of (\ref{sss})
\begin{eqnarray}\label{llll}
\cL_s & =&  
\cB + \cC +\cD
\nonumber\\
\nonumber\\
\cB & = &\pi^{-s} \bigg[\frac{\pi}{\tau}\bigg]^{\frac{1}{2}}
\frac{\Gamma[s-1/2]}{2} \sum_{n_1<0} z(n_1)^{\frac{1}{2}-s}
\nonumber\\
\nonumber\\
\cC & = & 2 \,\tau^{-\frac{s}{2}-\frac{1}{4}} \!\! \sum_{n_1<0;
\,p\geq 1}\!\! z(n_1)^{-\frac{s}{2}+\frac{1}{4}}
 p^{s-\frac{1}{2}} \cos\big[2 \pi c_1(n_1) p\big] \,
K_{s-\frac{1}{2}}\bigg(2\pi p  \,
\big({z(n_1)}/{\tau}\big)^{\frac{1}{2}}\bigg)
\nonumber\\
\nonumber\\
\cD & = &\pi^{-s} \sum_{n_1<0,\, m\geq  0}\frac{\Gamma[s+m]}{m!}
(-\tau)^m \zeta[-2 m, c_1(n_1)]\,{z(n_1)^{-s-m}}
\end{eqnarray}
The series in  $\cD$  is asymptotic \cite{elizalde2}. To compute
$\cB$ one considers the cases
 $\delta/(\tau U_2^2)< 1$,  $> 1$.
In the following we take $s=-2-\epsilon$.

\vspace{0.3cm} \noindent If $0\leq \delta/(\tau U_2^2)< 1$, one
uses for $\cB$ a binomial expansion of its term
$z(n_1)^{5/2+\epsilon}$, as in eqs.(\ref{zp}),
(\ref{expansionzeta}), and  the comments thereafter to isolate the
poles, to find

\begin{eqnarray}\label{bb1}
{\cal B}&=&
     \frac{1}{2}\bigg\{
\frac{\pi^3 \delta^3}{12\,\tau\, \vert U_2\vert}
\bigg[\frac{1}{\epsilon}\!+\ln\Big(4\pi\tau U_2^2
e^{\gamma+2\psi({1-\rho_1})} \Big)\bigg]+\pi^3 \delta^2 \vert
U_2\vert \Big[\frac{1}{12}-\frac{1}{2}\,\rho_1 (1-\rho_1)\Big]
\nonumber\\
\nonumber\\
 &&\qquad +
\frac{2\pi^3}{45} \tau^2 \vert U_2\vert^5
\bigg[\frac{1}{21}\!-\rho_1^2
  (1-5\rho_1^2 -\! 2 \rho_1^4+6 \rho_1^3)\bigg]\! -\frac{\pi^3 \delta
  \tau\vert U_2\vert^3}{90} \Big[1-30\rho_1^2 (1-\rho_1)^2\Big]
\nonumber\\
\nonumber\\
&&\qquad +\, \pi^{5/2} \, \tau^2 \vert U_2\vert^5 \sum_{p\geq 1}
\frac{\Gamma[p+1/2]}{(p+3)!} \bigg[\frac{-\delta}{\tau
    U_2^2}\bigg]^{p+3}\zeta[2p+1,1-{\rho_1}]\bigg\}+\cO(\epsilon)\qquad
\\
\nonumber
\end{eqnarray}
 If instead  $\delta/(\tau U_2^2)\geq 1$, one uses for
 $z(n_1)^{5/2+\epsilon}$ in $\cB$
the expansion  eq.(\ref{ff1}), to find

\begin{eqnarray}\label{bb2}
{\cal B}&=&
    \frac{1}{2}\bigg\{
\frac{\pi^3 \delta^3}{12 \tau \vert U_2
\vert}\Big[\frac{1}{\epsilon}+
    \ln\Big(\pi \delta e^{\gamma-\frac{11}{6}}\Big)\Big]
\!+
  2 \sqrt\tau U_2^2\delta^{\frac{3}{2}}\!\sum_{\tilde n_1\geq 1}
\frac{\cos[2\pi\tilde n_1\rho_1]}{\tilde n_1^3}\, K_3 \bigg(\!
2\pi \tilde n_1 \sqrt{\frac{\delta}{\tau U_2^2}}\!\bigg)
\nonumber\\
\nonumber\\
&&\qquad\qquad\quad + \frac{(\pi \delta)^{5/2}}{\sqrt
\tau}\sum_{m\geq 0}
    \frac{\Gamma[m-5/2]}{m!}
\Big(\frac{-\tau U_2^2}{\delta}\Big)^m\, \zeta[-2m,
1-\rho_1]\bigg\} +\cO(\epsilon)
\end{eqnarray}

\noindent For $\cC$ one  uses the definition of Bessel functions
$K_{-5/2}$ and of $\Li_\sigma$, to find

\begin{eqnarray}\label{cc}
\!\!\cC \!=\! \frac{\tau}{2} \!\!\sum_{n_1<0}\!\!\bigg[\! z(n_1)
\Li_3(e^{-2\pi \gamma(n_1)}) \!+\!\frac{3}{2\pi} (\tau
z(n_1))^{\frac{1}{2}} \Li_4(e^{-2\pi \gamma(n_1)}\!) \!+\!\frac{3
\tau }{4\pi^2}  \Li_5(e^{-2\pi\gamma(n_1)})\!+\!c.c.
\!\bigg]\!\!+\!\cO(\epsilon)\nonumber
\end{eqnarray}
\begin{eqnarray}\label{zzprime}
\gamma(n_1) & \equiv & \frac{1}{\sqrt\tau}(\delta+\tau U_2^2
(n_1+\rho_1)^2)^{1/2}- i(\rho_2-U_1(n_1+\rho_1))
\end{eqnarray}

\noindent Therefore if $0\leq \delta/(\tau U_2^2)\ll 1$ one has
from (\ref{llll}), (\ref{bb1}), (\ref{cc})

\begin{eqnarray}\label{lll}
\cL_{-(2+\epsilon)}&=& \cD+
 \frac{(\tau U_2)^2}{2}
\bigg\{ \frac{\pi^3 \delta^3}{12\,(\tau\, \vert U_2\vert )^3}
\frac{1}{\epsilon} +\frac{2\pi^3}{45}  \vert U_2\vert^3
\bigg[\frac{1}{21}\!-\rho_1^2
  (1-5\rho_1^2 -\! 2 \rho_1^4+6 \rho_1^3)\bigg]\!
\nonumber\\
\nonumber\\
&+& \sum_{n_1<0} \bigg[ (n_1+\rho_1)^2 \, \Li_3(\sigma_{n_1})
\!+\!\frac{3 \,\vert n_1+\rho_1\vert }{2\pi \vert U_2\vert}\,
\Li_4(\sigma_{n_1}) \!+\!\frac{3\, }{4\pi^2 U_2^2}  \,
\Li_5(\sigma_{n_1})\!+\!c.c. \bigg]\bigg\}
\nonumber\\
\nonumber\\
&&\sigma_{n_1<0}=e^{-2 i \pi
  \big({\rho_2}-U_1 \big\vert {\rho_1}+n_1\big\vert\big)}
\,\, e^{2\pi\vert U_2\vert (\rho_1+n_1)},\qquad 0\leq \rho_{1,2}<1
\end{eqnarray}

\vspace{0.3cm} \noindent Eq.(\ref{vorb2}) in the text  is then
$V_\cO=-\mu^4/(2\pi)^{4-\epsilon} \cL_{-(2-\frac{\epsilon}{2})}
\big(\delta\!\ra\! {m^2}/{\mu^2},\tau\!\ra\! (2\pi)^2/(\mu^2\cA
U_2)\big)$.

\noindent It remains to evaluate  the  series of $\cD$ in
(\ref{llll}) in a form  amenable to numerical evaluation. For
this, its factor $z(n_1)$ under the sum  can be expanded  in
$\delta/(\tau U_2^2)$ if $0\leq \delta/(\tau U_2^2)<1$ by using
the binomial expansion  eq.(\ref{zp}). If $\delta/(\tau U_2^2)>1$
one uses instead the asymptotic expansion of eq.(\ref{ff1}). For
our purposes $s=-(2+\epsilon)$, and then

\begin{equation}\label{dfdf}
\cD  =  \pi^{2+\epsilon}\!\!\!\!\! \sum_{n_1<0,\, m\geq  0}\!\!\!
 \Gamma[m-2-\epsilon]
\frac{(-\tau)^m}{m!} \zeta[-2 m, \rho_2-U_1 (n_1+\rho_1)]\,
\Big[\delta+\tau U_2^2(n_1+\rho_1)^2\Big]^{2-m+\epsilon}
\end{equation}
Eq.(\ref{ddefinition}) in the text is  $\tilde \cD \equiv 2 \cA^2
\,\mu^4/(2\pi)^{4-\epsilon}
\,\cD\Big(\epsilon\!\ra\!-{\epsilon}/{2},\tau\!\ra\!
(2\pi)^2/(\mu^2 U_2 \cA),
 \delta\!\ra\! m^2/\mu^2\Big)$.

\vspace{0.2cm} \noindent $\bullet$ {\bf  Case of $\cT_2/Z_4$
orbifold: $    $} In the following we restrict the calculation of
$\cD$ to
  $\cT_2/Z_4$, when $U_1=0$, $U_2=1$.
If so, the  argument of zeta function in $\cD$ does not have a
$n_1$
 dependence and the  sums over $n_1$ and $m$
can be easily performed. (For other orbifolds $U_1\not=0$ further
evaluation of $\cD$ is more tedious but very similar).

\vspace{0.1cm} \noindent {\bf (a).} For $0\leq \delta/\tau <1$,
after a binomial expansion (\ref{zp}) of last bracket in
(\ref{dfdf}),  $\cD$ becomes

\begin{eqnarray}\label{dd}
\cD& = &
 (\pi\tau)^{2+\epsilon}\!\!\!\!\!
\sum_{n_1<0,\, m\geq  0}\!\!\!
 \Gamma[m-2-\epsilon]
\frac{(-1)^m}{m!} \zeta[-2 m, \rho_2]\,
\Big[\delta/\tau+(n_1+\rho_1)^2\Big]^{2-m+\epsilon}
\nonumber\\
\nonumber\\
&=& (\pi\tau)^{2+\epsilon}\!\!\sum_{m\geq 0, p\geq 0}
\Big[\frac{\delta}{\tau}\Big]^p\, (-1)^{m+p}
\frac{\Gamma[p+m-2-\epsilon]}{m!\, p!} \zeta[-2 m, \rho_2]\,
\zeta[2 p+2m-4-2\epsilon,1-\rho_1]
\nonumber\\
\nonumber\\
&=& \frac{(\pi\tau)^2}{2 \epsilon} \bigg\{\zeta[-4,\rho_2]\,
\zeta[0,\rho_1]+\zeta[-4,\rho_1]\, \zeta[0,\rho_2]+2
\zeta[-2,\rho_1]\,\zeta[-2,\rho_2]\bigg\}
\nonumber\\
\nonumber\\
&+& \frac{\pi^2\delta\tau}{\epsilon} \bigg\{\zeta[0,\rho_1]
\,\zeta[-2,\rho_2]+\zeta[-2,\rho_1]\,\zeta[0,\rho_2]\bigg\}
+\frac{\pi^2 \delta^2}{2\epsilon}
\zeta[0,\rho_1]\,\zeta[0,\rho_2]+\cD_f+\cO(\epsilon)\\
\nonumber
\end{eqnarray}
used in (\ref{coeffs}), (\ref{vz4}) with $\tilde \cD \equiv 2
\mu^4 L^4/(2\pi)^{4-\epsilon}
\,\cD\big(\epsilon\!\ra\!-\epsilon/2,\tau\!\ra\! (2\pi)^2/(\mu
L)^2,
 \delta\!\ra\! m^2/\mu^2\big)$.\newline
$\cD_f$ in eq.(\ref{dd}) is the finite $\cO(\epsilon^0)$  part:

\begin{eqnarray}\label{rrr}
\cD_f\!\!&=&\!\! \frac{(\pi\tau)^2}{2}\!
\bigg[2\zeta[-2,\rho_1]\zeta[-2,\rho_2]\ln(\tau \pi e^{\gamma-1})
\!+\! \zeta[0,\rho_2]\Big(\zeta[-4,\rho_1]\ln(\pi\tau
e^{\gamma-\frac{3}{2}}) \!+ \! 2  \zeta'[-4,1-\rho_1]\Big)
\nonumber\\
\nonumber\\
&+& 4 \zeta[-2,\rho_2] \zeta'[-2,1-\rho_1] +\zeta[-4,\rho_2]
\Big(\zeta[0,\rho_1]\ln(\pi\tau e^\gamma)\!+\!
2\zeta'[0,1-\rho_1]\Big) \bigg]
\nonumber\\
\nonumber\\
&+& \pi^2 \delta \tau \bigg[\zeta[0,\rho_2]\Big(
\zeta[-2,\rho_1]\ln(\pi\tau
e^{\gamma-1})+2\zeta'[-2,1-\rho_1]\Big) +
\zeta[-2,\rho_2]\Big(\zeta[0,\rho_1]\ln(\pi\tau e^{\gamma})
\nonumber\\
\nonumber\\
&+& 2\zeta'[0,1-\rho_1]\Big)\bigg] + \frac{(\pi \delta)^2}{2}
\zeta[0,\rho_2]\Big(\zeta[0,\rho_1] \ln(\pi\,\tau e^\gamma)
+2\zeta'[0,1-\rho_1]\Big)
\nonumber\\
\nonumber\\
&+&\!\!\!\!\! \sum_{p\geq 0, m\geq 0, p+m\geq 3}\!\! \pi^2\tau^2
\Big(\frac{\delta}{\tau}\Big)^p (-1)^{p+m}
\frac{\Gamma[p+m-2]}{p!\,m!}
\zeta[-2 m,\rho_2]\,\zeta[2 p+2 m-4,1-\rho_1]\\
\nonumber
\end{eqnarray}
This was used in (\ref{vorbz4}), (\ref{finitep}) with $\tilde
\cD_f \equiv 2 \mu^4 L^4/(2\pi)^4 \,\cD_f\big(\tau\!\ra\!
(2\pi)^2/(\mu L)^2,
 \delta\!\ra\! m^2/\mu^2\big)$, after neglecting any $m L\ll 1$
dependence.

\vspace{0.5cm} \noindent {\bf (b).} In the case when
$\delta/\tau>1$ one uses  in $\cD$ of (\ref{dfdf}) or the first
line in (\ref{dd}),   the asymptotic expansion eq.(\ref{ff1}). The
results shows that the divergent part of $\cD$ is identical to
that in the last two lines in (\ref{dd}), while the value of
$\cD_f$ ($\cO(\epsilon^0)$) in  (\ref{dd}) has now the expression

\begin{eqnarray}\label{qqq}
\cD_f & = & (\pi \tau)^2\!\!\!\!\!\!\! \!\! \sum_{m\geq 0, k\geq
0, k+m\geq 3}\!\!\!\!\!\!\!\!\!\!
 (-1)^{m+k}\,  \zeta[-2k, 1-\rho_1]\, \zeta[-2 m,\rho_2]
\, \frac{\Gamma[k+m-2]}{m!\, k!}
\, \Big( \frac{\delta}{\tau} \Big)^{2-m-k}\\
 & +&
 \sum_{m\geq 0}
 \frac{\pi^{\frac{5}{2}} \tau^2}{2} \zeta[-2 m,\rho_2] \frac{(-1)^m}{m!}
 \Big(\frac{\delta}{\tau}\Big)^{\frac{5}{2}-m}
 \Gamma[m-5/2]\nonumber\\
 \nonumber\\
 &+&
\sum_{m\geq 0} \frac{2\tau^2 (-\pi)^m}{m!}
\Big(\frac{\delta}{\tau}\Big)^{\frac{5}{4}-\frac{m}{2}}\zeta[-2m,\rho_2]
\sum_{p\geq 1}\cos[2\pi p\rho_1]p^{m-\frac{5}{2}}
\,K_{m-\frac{5}{2}} \Big(2\pi p \sqrt{\frac{\delta}{\tau}}\Big)
\nonumber\\
\nonumber\\
&+& \frac{\pi^2\tau^2}{2}\bigg[ \zeta[-4,\rho_2] \zeta[0,\rho_1]+
\zeta[-4,\rho_1] \zeta[0,\rho_2]+
2\zeta[-2,\rho_1]\zeta[-2,\rho_2]\bigg]\ln(\pi\delta e^\gamma)
\nonumber\\
\nonumber\\
+&&\!\!\!\!\!\! \pi^2 \tau \delta(\zeta[-2,\rho_1]
\zeta[0,\rho_2]+ \zeta[-2,\rho_2]\, \zeta[0,\rho_1])\ln(\pi\delta
e^{\gamma-1}) +
\frac{\pi^2\delta^2}{2}\zeta[0,\rho_1]\,\zeta[0,\rho_2]
\ln(\pi\delta e^{\gamma-\frac{3}{2}})\nonumber\\
\nonumber
\end{eqnarray}

\noindent To conclude, if $\delta/\tau \geq 1$, the value of
$\cL_{-(2+\epsilon)}$ is given by
\begin{eqnarray}\label{ptr}
\cL_{-(2+\epsilon)}&=& \cB+\cC+\cD_f
\nonumber\\
\nonumber\\
&+& \frac{(\pi\tau)^2}{2 \epsilon} \bigg\{\zeta[-4,\rho_2]\,
\zeta[0,\rho_1]+\zeta[-4,\rho_1]\, \zeta[0,\rho_2]+2
\zeta[-2,\rho_1]\,\zeta[-2,\rho_2]\bigg\}
\nonumber\\
\nonumber\\
&+& \frac{\pi^2\delta\tau}{\epsilon} \bigg\{\zeta[0,\rho_1]
\,\zeta[-2,\rho_2]+\zeta[-2,\rho_1]\,\zeta[0,\rho_2]\bigg\} +
\frac{\pi^2 \delta^2}{2\epsilon} \zeta[0,\rho_1]\,\zeta[0,\rho_2]
+ \cO(\epsilon)\qquad\quad
\end{eqnarray}
$\cB$ is given in eq.(\ref{bb2}),  $\cC$ in eq.(\ref{cc}), while
$\cD_f$ is that of  eq.(\ref{qqq}).

Eq.(\ref{ptr}) concludes the calculation of $\cL_{-(2+\epsilon)}$
for $\cT_2/Z_4$ for $\delta/\tau\geq 1$. Eq.(\ref{lll}) with
(\ref{dd}), (\ref{rrr}), gives $\cL_{-(2+\epsilon)}$ for
$0<\delta/\tau<1$ again for $\cT_2/Z_4$. Eq.(\ref{lll}) with
(\ref{bb1}), (\ref{bb2}), (\ref{cc}), (\ref{dfdf}) give
$\cL_{-(2+\epsilon)}$ for any $U$.

\vspace{0.7cm} With the expression (\ref{ptr})  for $\cL$,
eq.(\ref{vlargeml})  in the text  is then
\begin{eqnarray}\label{erwsdfgv}
V_\cO=-\frac{\mu^4}{(2\pi)^{4-\epsilon}}
\cL_{-(2-\frac{\epsilon}{2})} \Big(\delta\!\ra\!
\frac{m^2}{\mu^2},\tau\!\ra\! \frac{(2\pi)^2}{ (\mu L)^2}\Big).
\end{eqnarray}

\newpage

\def\theequation{\thesubsection-\arabic{equation}}
\def\thesubsection{D}
\setcounter{equation}{0}
\subsection{.\,\, Orbifolds, Fixed points and discrete Wilson lines.}
\label{appendixF} \setcounter{equation}{0}

The lattice of $T_2/Z_N$ orbifolds is generated by $(1,\xi)$ with
$\xi=i$ for $Z_2$, $Z_4$ and $\xi=e^{2 i \pi/3}$ for $Z_3$, $Z_6$.
The group $Z_N$ of discrete rotations
 has $N$ elements $P^n_N$, $0\leq n\!<\!N\!-\!1$ with
$P^N_N=1$. Their  fixed points are
\begin{eqnarray}
Z_2:&& \qquad \qquad P_2:\qquad  z_{f.p.}=0,\,\, 1/2,
\,\,\xi/2,\,\, (1+\xi)/2, \qquad\qquad \quad \xi=i
\nonumber\\
\nonumber\\
Z_3: && \qquad\,\, P_3, P_3^2:\qquad z_{f.p.}=0,\,\, (2+\xi)/3,
\,\,(1+2 \xi)/3, \qquad\qquad \xi=e^{2 i \pi/3}
\nonumber\\
\nonumber\\
Z_4: &&  \qquad\,\; P_4, P_4^3:\qquad  z_{f.p.}=0, \,\,(1+\xi)/2,
 \qquad\qquad \qquad\qquad  \,\quad \xi=i\nonumber\\
      &&  \qquad\qquad P^2_4:\qquad  z_{f.p.}=0,\,\,
 (1+\xi)/2,\,\, \xi/2,\,\, 1/2.
\nonumber\\
\nonumber\\
Z_6: && \qquad \,\; P_6, P_6^5:\qquad
z_{f.p.}=0.\hspace{5.5cm}\xi=e^{2i\pi/3}
\nonumber\\
 && \qquad \,\, P_6^2, P_6^4:\qquad z_{f.p.}=0,\,\,(2+\xi)/3,\,\, (1+2\xi)/3,
\nonumber\\
&& \qquad \qquad P_6^3:\qquad z_{f.p.}=0,\,\, 1/2,\,\,\xi/2,\,\,
(1+\xi)/2.
 \end{eqnarray}
The usual orbifold action ($g$) and that of
 Wilson lines ($T_{1,2}$)  are given by
\begin{eqnarray}
\Phi^g(z+1)&=&T_1\, \Phi^g(z),\nonumber\\
\Phi^g(z+\tau)&=&T_2\, \Phi^g(z) \nonumber\\
\Phi^g(\tau z)&=& g \,\Phi^g(z),\quad {\rm with}\qquad \tau\equiv
e^{2 \,i\,\pi/N}
\end{eqnarray}
One has that
\begin{eqnarray}
\Phi^g(z+\tau)=\Phi^g(\tau (\tau^{-1} z+1))=g\, T_1\,
\Phi^g(\tau^{-1}z)=g \,T_1 \, g^{N-1}\, \Phi^g(z)
\end{eqnarray}
Using the definition of $T_2$, then
\begin{eqnarray}
T_2 \,\,g \, =\, g \,\,T_1
\end{eqnarray}
One can further assume that the orbifold action $g$ and the Wilson
lines $T_i$ commute, then $T_1=T_2=T$, and  with  $T_i\equiv e^{2
i\pi \,\rho_i}$ one
 finds (modulo ${\bf Z}$) that  $\rho_1=\rho_2=\rho$ for any $T_2/Z_N$.

\vspace{0.5cm} \noindent {\bf The case of $Z_3$ orbifolds:}
($\xi=e^{2 i\pi/3}$)
\begin{eqnarray}
\Phi^g(\tau^2 (z+1))& = & g^2\,T_1\, \Phi^g(z)
\nonumber\\[10pt]
\Phi^g(\tau^2 (z+1)) & = & T_1^{-1} \,\Phi^g(\tau^2
(z+1)+1)=T_1^{-1} T_2^{-1}\,\Phi^g(\tau^2 z)=T_1^{-1} T_2^{-1}
\,g^2\, \Phi^g(z)
\end{eqnarray}
A  solution to this  is
\begin{eqnarray}
g^2\,T_1=T_1^{-1} T_2^{-1}\,g^2
\end{eqnarray}
or, assuming $[g,T_i]=0$ giving  $T_1=T_2=T$
\begin{eqnarray}
T^3=1,\qquad \Rightarrow \qquad \rho=0,\,1/3,\,2/3.
\end{eqnarray}
where $T=\exp(2 i\pi \rho)$. Further, for the fixed points
\begin{eqnarray}
\!\!\!\!\!\! a). \quad if \quad z_{f.p.}&=&0,
\qquad\qquad\quad\,\, \Rightarrow  \quad \Phi^g(0)=\Phi^g(\tau\,
0)=g\,\, \Phi^g(0)
\nonumber\\
b). \quad if \quad z_{f.p.}&=&(2+\xi)/3, \qquad \Rightarrow \quad
T\,g\,\, \Phi^g(z_{f.p.})\,= \Phi^g(z_{f.p.})
\nonumber\\
c). \quad if \quad z_{f.p.}&=& (1+2\,\xi)/3,\quad\,\,\,
\Rightarrow \quad T^2 \,g \,\Phi^g(z_{f.p.})= \Phi^g(z_{f.p.})= T
\,g^2\, \Phi^g(z_{f.p.})
\end{eqnarray}
These are additional conditions which must be respected by the
Wilson lines $T$, orbifold projections $g$ and fields $\Phi^g$
 at the fixed points. The conditions can be respected by
suitable relative choices for $T$, $g$,  or trivially by requiring
the fields vanish at these fixed points.

\vspace{0.5cm} \noindent {\bf The case of $Z_4$ orbifolds:}
($\xi=i$)
\begin{eqnarray}
\Phi^g(\tau^2 (z+1))& =& T_1^{-1} \Phi^g(\tau^2
z+\tau^2+1)=T_1^{-1} g^2\Phi^g(z)
\nonumber\\[10pt]
\Phi^g(\tau^2 (z+1)) & =& g^2\,T_1\,\Phi^g(z)
\end{eqnarray}
which gives
\begin{eqnarray}
g^2\, T_1= T^{-1}_1\, g^2
\end{eqnarray}
or, assuming $[g,T_i]=0$ giving $T_1=T_2=T$ one has
\begin{eqnarray}
T^2=1,\qquad \Rightarrow \qquad \rho=0,1/2.
\end{eqnarray}
where $T=\exp(2 i\pi \rho)$. Further, for the fixed points
\begin{eqnarray}
a).\quad if \quad z_{f.p.}&=&0,\qquad \qquad\,\,  \Rightarrow
\qquad \Phi^g(0)=\Phi^g(\tau\, 0)=g\,\, \Phi^g(0)
\nonumber\\
b). \quad if \quad z_{f.p.}&=& (1+\xi)/2,\quad\Rightarrow \qquad
T\,g\,\, \Phi^g(z_{f.p.})\,= \Phi^g(z_{f.p.})
\nonumber\\
c).\quad if \quad  z_{f.p.}&=& 1/2, \qquad\quad\,\,\, \Rightarrow
\qquad T\,g^2 \,\Phi^g(z_{f.p.})= T\,\Phi^g(-1/2)=\Phi^g(z_{f.p.})
\nonumber\\
d). \quad if \quad  z_{f.p.}&=& \xi/2,\,\,\,\qquad\quad
\Rightarrow \qquad
T\,g^2\,\Phi^g(z_{f.p.})=g\,T\,\Phi^g(\xi^2/2)=\Phi^g(z_{f.p.})
\end{eqnarray}
Similar to the $Z_3$ case, the conditions can be respected by
suitable choices for $T$, $g$,  or trivially by requiring the
fields vanish at these fixed points.

\vspace{0.5cm} \noindent {\bf The case of $Z_6$ orbifolds:}
\begin{eqnarray}
\Phi^g(\tau^3 (z+1)) & =& g^3\,T_1\,\Phi^g(z)
\nonumber\\[10pt]
\Phi^g(\tau^3 (z+1))& =& T_1^{-1} \, \Phi^g(\tau^3
z+\tau^3+1)=T_1^{-1}\, g^3\Phi^g(z)
\end{eqnarray}
since $\tau^3+1=0$. One solution is
\begin{eqnarray}
g^3\, T_1= T^{-1}_1\, g^3
\end{eqnarray}
Assuming $[g,T_1]=0$ which gives $T_1=T_2=T$ one has
\begin{eqnarray}\label{s1}
T^2=1,\quad \Rightarrow \quad \rho=0,1/2.
\end{eqnarray}
where $T\equiv \exp(2 i \pi \rho)$. Further,
\begin{eqnarray}
\Phi^g(\tau^2 (z+1)) & =& g^2\,T_1\,\Phi^g(z)
\nonumber\\[10pt]
\Phi^g(\tau^2 (z+1))& =& T_1^{-1} \, \Phi^g(\tau^2
z+\tau^2+1)=T_1^{-1} T_2\, g^2\Phi^g(z)
\end{eqnarray}
since $\tau^2-\tau+1=0$. One solution is
\begin{eqnarray}
g^2\, T_1= T^{-1}_1\,T_2 \, g^2
\end{eqnarray}
With $[g,T_i]=0$ giving $T_1=T_2=T$ one has
\begin{eqnarray}\label{s2}
T=1\,\qquad\Rightarrow\qquad \rho=0.
\end{eqnarray}
where $T\equiv \exp(2 i \pi \rho)$. Thus, if $[g,T_i]=0$,
 one concludes from (\ref{s1}), (\ref{s2}) that  $\rho=0$.
 Further relations at the fixed points
exist, which can be found as in  the case of
$\cT_2/Z_{3,4}$.

\def\theequation{\thesubsection-\arabic{equation}}
\def\thesubsection{E}
\setcounter{equation}{0}
\subsection{. \, Mathematical Formulae and Conventions.}
\label{appendixE} \setcounter{equation}{0} We used the Poisson
re-summation formula
\begin{equation}\label{p_resumation}
\sum_{n\in Z} e^{-\pi A (n+\sigma)^2}=\frac{1}{\sqrt A}
\sum_{\tilde n\in Z} e^{-\pi A^{-1} \tilde n^2+2 i \pi \tilde n
\sigma}
\end{equation}
The integral representation of Bessel Function $K_\nu$ \cite{gr}
\begin{equation}\label{bessel1}
\int_{0}^{\infty} \! dx\, x^{\nu-1} e^{- b x^p- a
x^{-p}}=\frac{2}{p}\, \bigg[\frac{a}{b} \bigg]^{\frac{\nu}{2 p}}
K_{\frac{\nu}{p}}(2 \sqrt{a \, b}),\quad Re (b),\, Re (a)>0
\end{equation}
with
\begin{eqnarray}\label{k52}
K_{-\frac{5}{2}}(z)& =&\sqrt{\frac{\pi}{2z}}
e^{-z}\bigg[1+\frac{3}{z^2}+\frac{3}{z^2} \bigg]
\nonumber\\
\nonumber\\
K_3(z\gg 1)&=& e^{-z} \sqrt\frac{\pi}{2
z}\bigg[1+\frac{35}{8}\frac{1}{z}
+\frac{945}{128}\frac{1}{z^2}+\frac{3465}{1024}\frac{1}{z^3}+\cdots\bigg]
\end{eqnarray}
The definition of PolyLogarithm
\begin{eqnarray}\label{plg}
Li_\sigma(x)=\sum_{n\geq 1} \frac{x^n}{n^\sigma}
\end{eqnarray}

\end{document}